\documentclass[12pt,showkeys,preprintnumbers,superscriptaddress,amsmath,amssymb,nofootinbib]{revtex4-1}

\usepackage{graphicx}
\usepackage{dcolumn}
\usepackage{bm}
\usepackage{amssymb}
\usepackage{amsmath}
\usepackage{epsfig}    
\usepackage{color}
\usepackage{slashed}
\usepackage{makecell}
\usepackage{float} 
\usepackage[hidelinks]{hyperref}
\usepackage{tikz}
\usetikzlibrary{decorations.pathmorphing}
\usepackage{placeins}
\FloatBarrier
\begin{document}

\title{Double SM-like Higgs production at future  $ e^+ e^- $ colliders in the 3-Higgs Doublet Model under the $ S_{3} $ symmetry }

\author{Emine Yildirim}
\email{emineyildirim@ktu.edu.tr}
\affiliation{Department of Physics, Karadeniz Technical University, Trabzon, TR61080, Türkiye}

\begin{abstract}
\noindent

In this paper, we present the production of  double Standard Model (SM)-like Higgs $ (h) $  at the future electron positron colliders   within the framework of the $ S_{3} $ model with three Higgs doublets (S3-3H) and no CP violation  to describe beyond the SM Higgs Physics. We focus first on the numerically allowed parameter space of the model, taking into account  theoretical bounds from perturbative unitarity and vacuum stability, as well as by data at the Large Hadron Collider (LHC) and the Tevatron. Double Higgs production  in the S3-3H model can deviate from  the SM  predictions up to a few orders of magnitude in both the  $ e^+e^- \rightarrow hhZ $ and  the $ e^+ e^- \rightarrow  h h\nu_{e} \bar{\nu}_{e}$ channels. Thus, our findings indicate that the S3-3H  model can lead  to sizable deviations from the SM predictions of Higgs production at future  $ e^+ e^- $ colliders. These results highlight the importance of  studies at such   colliders for  searching for physics beyond the SM.
\end{abstract}


\maketitle
\newpage

\newpage
\section{INTRODUCTION}

The discovery of a Higgs boson with a mass near 125 GeV is a major milestone in particle physics,  validating the mechanism responsible for electroweak symmetry breaking (EWSB) within the SM\cite{1,2}. In spite of this achievement, our understanding of the Higgs sector remains incomplete. In particular, the experiments have  yet not determined the detailed structure of the scalar potential and the dynamics governing Higgs interactions  fully. While current data shows that the discovered scalar closely resembles the SM Higgs boson, its couplings, particularly the self-couplings, have not yet been tested with sufficient precision.
The trilinear Higgs self-coupling is of particular importance, as it directly tests the structure of the scalar potential and leads to fundamental questions such as vacuum stability and the nature of the electroweak phase transition. Furthermore, the scalar sector leads to the hierarchy problem in the absence of a symmetry that stabilizes the Higgs mass. These issues motivate the study of scenarios beyond the SM, many of which predict additional scalar states and modified Higgs couplings. High-precision measurements at current and future collider experiments are required to test the SM and investigate the possibility of an extended Higgs sector.

Extensions of the SM scalar sector offer a well-motivated framework to test possible modifications in Higgs interactions and to explore Higgs physics beyond the SM. Multi-Higgs doublet models based on discrete symmetries provide a rich phenomenology with a predictive nature. In this work, we will focus on the S3-3H model. The  model   is built by applying the permutation symmetry among three scalar doublets, which can be split into one singlet and one doublet representation under the S$_3$ group.  The S3-3H model has been studied in different frameworks \cite{GomezBock:2021,66,65,67,69,70,71,72}. As well as the S$_3$ symmetry, a residual $Z_2$ symmetry can appear after electroweak symmetry breaking, leading to definite $ Z_{2} $ parity assignments for the scalar fields. \cite{GomezBock:2021,66}. The S3-3H model offers nine physical bosons, one of which acts as a SM-like Higgs boson. The Higgs couplings to gauge bosons and Higgs self-coupling in  the S3-3H model differ from those in the SM. The S3-3H model  is quite convenient for studying double Higgs production processes,  due to these properties. Furthermore, the existence of additional heavy scalars  can cause  resonant contributions that significantly enhance the production cross section, leading to a distinctive characteristic of an extended Higgs sector. In this paper, we will investigate how an extended Higgs sector, namely, the  S3-3H model in Ref.~\cite{GomezBock:2021} can affect  the production cross sections at future $ e^+e^- $ colliders.  Recently there has been increased interest in the phenomenology of extended Higgs sectors. Many studies of extended Higgs sectors have focused on  theoretical frameworks and their implications for collider phenomenology, including precision Higgs measurements, exotic scalar decays, flavor observables, and searches for additional neutral and charged Higgs bosons at current and future collider facilities \citep{ex1,ex2,ex6,ex3,ex4, ex5, ex7}. These advancements encourage detailed studies of specific models, such as the S3-3H model investigated in this paper.

  $ e^+ e^- $ colliders offer a  clean environment for  precision studies of the Higgs sector and allow more precise measurements of Higgs couplings to gauge bosons, fermions, and to itself as compared to hadron colliders due to the absence of QCD backgrounds. Several relevant works are presented in Ref.~\cite{Li:2017daq,18,19,43,44,20,21,22,23,24,25,26,27,28,29,30,31,32,33,34,41,45,100,101}. Double Higgs $ (h) $ production modes, including $e^+e^- \rightarrow hhZ$ and $ e^+ e^- \rightarrow  h h\nu_{e} \bar{\nu}_{e}$, offer the possibility to measure the trilinear Higgs self-coupling, which is  important for building the scalar potential.   Future high-energy $e^+e^-$ colliders, such as CLIC, can measure the Higgs self-coupling with a precision at the level of $\mathcal{O}(10\%)$ at multi-TeV energies~\cite{Lukic:2016}. The channel  $ e^+ e^- \rightarrow  h h\nu_{e} \bar{\nu}_{e}$   becomes dominant and increases the sensitivity to both Higgs self-interactions and Higgs-gauge interactions at the center-of-mass energies  above $\sqrt{s} \sim 500$ GeV. In addition, one can access the coupling $ g_{Hhh} $ of the  CP-even heavy Higgs boson  in the $ e^+ e^- \rightarrow  h h\nu_{e} \bar{\nu}_{e}$  process through its propagators. In short, precision measurements of the Higgs boson couplings are a useful way to identify the nature of EWSB.  Deviations from the SM predictions may indicate the presence of an extended Higgs sector. One can access these deviations indirectly through studies of pair production cross sections. In this paper, our main goal is to  focus on the production modes of  $e^+e^- \rightarrow hhZ$ and $ e^+ e^- \rightarrow  h h\nu_{e} \bar{\nu}_{e}$. Our calculation of  $ \sigma^{S3-3H}/\sigma^{SM} $ shows considerable deviations from unity in both channels.

The rest of the paper is planned as follows. In Section~II, we summarize the essential features in the scalar  and gauge sectors of the  S3-3H model and list the relevant couplings used in the phenomenology section. 
In Section~III, we discuss both theoretical and experimental constraints on the parameter space of the model.  Section~IV presents the $ e^{+}e^{-} $ phenomenology of double Higgs production modes.
The  conclusions of the paper are drawn in Section~V. 
In Appendix A and B, relevant coefficients used in  the phenomenological study are presented. 

\section{THE S3-3H MODEL FRAMEWORK}
The S3-3H model  has been discussed in detail  in the original work Ref.~\cite{GomezBock:2021}.  We consider the  S3-3H model, following the formulation given in Ref.~\cite{GomezBock:2021}. 
In this model, three  $SU(2)_L$ electroweak (EW) Higgs  scalar doublets are introduced, where two of them, $(H_1,H_2)$, transform as an $S_3$ doublet, while the third field, $H_S$, transforms as the symmetric singlet representation of $S_3$.  We adopt the scalar and kinetic sectors  of the model from Ref.~\cite{GomezBock:2021} and follow its notations. However, we do not reproduce the full expression here.

The  general scalar    potential invariant under
$SU(2)_L \times U(1)_Y \times S_3$ is  given in Ref.~\cite{GomezBock:2021} as: 
\begin{equation}
\begin{aligned}
V &= \mu_1^2 \left( H_1^\dagger H_1 + H_2^\dagger H_2 \right)
+ \mu_0^2 \left( H_s^\dagger H_s \right)
+ \frac{a}{2} \left( H_1^\dagger H_s \right)^2
+ b \left( H_s^\dagger H_s \right)\left( H_1^\dagger H_1 + H_2^\dagger H_2 \right) \\
&\quad + \frac{c}{2} \left( H_1^\dagger H_1 + H_2^\dagger H_2 \right)^2
+ \frac{d}{2} \left( H_1^\dagger H_2 - H_2^\dagger H_1 \right)^2
+ e f_{ijk} \left( H_s^\dagger H_i \right)\left( H_j^\dagger H_k \right) + \text{h.c.} \\
&\quad + f \Big\{
\left( H_s^\dagger H_1 \right)\left( H_1^\dagger H_s \right)
+ \left( H_s^\dagger H_2 \right)\left( H_2^\dagger H_s \right)
\Big\} \\
&\quad + \frac{g}{2} \Big\{
\left( H_1^\dagger H_1 - H_2^\dagger H_2 \right)^2
+ \left( H_1^\dagger H_2 + H_2^\dagger H_1 \right)^2
\Big\} \\
&\quad + \frac{h}{2} \Big\{
\left( H_s^\dagger H_1 \right)\left( H_s^\dagger H_1 \right)
+ \left( H_s^\dagger H_2 \right) \left( H_s^\dagger H_2 \right)
+ \left( H_1^\dagger H_s \right)\left( H_1^\dagger H_s \right)
+ \left( H_2^\dagger H_s \right) \left( H_2^\dagger H_s \right)
\Big\};
\end{aligned}
\end{equation}
where $f_{112} = f_{121} = f_{211} = -f_{222} = 1$.

Higgs doublets are given  in terms of complex fields as follows: 
\begin{equation}
H_1 = \frac{1}{\sqrt{2}}
\begin{pmatrix}
\phi_1 + i \phi_4 \\
\phi_7 + i \phi_{10}
\end{pmatrix},
\quad
H_2 = \frac{1}{\sqrt{2}}
\begin{pmatrix}
\phi_2 + i \phi_5 \\
\phi_8 + i \phi_{11}
\end{pmatrix},
\quad
H_s = \frac{1}{\sqrt{2}}
\begin{pmatrix}
\phi_3 + i \phi_6 \\
\phi_9 + i \phi_{12}
\end{pmatrix}. \label{vev}
\end{equation}

After spontaneous symmetry breaking in the absence of CP violation, only the real components of each scalar doublet can obtain vacuum expectation values (VEVs). The VEVs are parameterized in terms of the field components of $H_1$, $H_2$, and $H_s$ defined in Eq.~(\ref{vev})  as
\begin{equation}
\langle \phi_7 \rangle = v_1,\;
\langle \phi_8 \rangle = v_2,\;
\langle \phi_9 \rangle = v_S,\;
\langle \phi_i \rangle = 0,\quad i \neq 7,8,9.
\end{equation}

The SM vev is written as 
\begin{equation}
v = \sqrt{v_1^2 + v_2^2 + v_S^2}.
\end{equation}
The VEVs are parametrized in spherical coordinates as
\begin{equation}
v_1 = v \cos\varphi \sin\theta, \quad
v_2 = v \sin\varphi \sin\theta, \quad
v_S = v \cos\theta,
\end{equation}
with the relations
\begin{equation}
\tan\varphi = \frac{v_2}{v_1}, \quad
\tan\theta = \frac{v_2}{v_S \sin\varphi}.
\end{equation}

The minimization conditions for the scalar potential give:
\begin{equation}
v_{1}^{2}=3v_{2}^{2}
\end{equation}
 The S3-3H model predicts the vacuum alignment $v_1 = \sqrt{3}\,v_2$ by fixing the angle $\phi$ to $\pi/6$~\cite{66,GomezBock:2021}. Therefore  $\phi=\pi/6$  is  not treated as an independent input parameter in the numerical analysis. After EWSB, this choice gives rise to a residual $Z_2$ symmetry, associated with one of the $Z_2$ subgroups of $S_3$. It can be interpreted as a reflection symmetry along one axis of an equilateral triangle. We note that the general $S_3$-symmetric scalar potential may lead to other vacuum solutions. Consequently, the numerical results discussed in this work are specific to the $Z_2$-preserving vacuum realization considered here and should not be regarded as generic predictions of all possible $S_3$-3HDM vacuum configurations.
 
\subsection{Higgs Basis and Mass Eigenstates}

In multi-Higgs models, it is convenient to work in the so-called Higgs basis, in which only one scalar field acquires the full vacuum expectation value, while the other fields are perpendicular to it. In this basis, the scalar fields are rotated as
\begin{equation}
\begin{pmatrix}
\phi_{\text{vev}} \\
\psi_1 \\
\psi_2
\end{pmatrix}
=
R_A^T
\begin{pmatrix}
H_1 \\
H_2 \\
H_s
\end{pmatrix},
\end{equation}
where the rotation matrix is given by
\begin{equation}
R_A^T =
\begin{pmatrix}
\sin\theta \cos\varphi & \sin\theta \sin\varphi & \cos\theta \\
-\sin\varphi & \cos\varphi & 0 \\
-\cos\theta \cos\varphi & -\cos\theta \sin\varphi & \sin\theta
\end{pmatrix}.
\end{equation}

The rotation matrix can be written as
\begin{equation}
R_A^T =
\begin{pmatrix}
\frac{\sqrt{3} v_2}{v} & \frac{v_2}{v} & \frac{v_S}{v} \\
-\frac{1}{2} & \frac{\sqrt{3}}{2} & 0 \\
-\frac{\sqrt{3} v_S}{2v} & -\frac{v_S}{2v} & \frac{2 v_2}{v}
\end{pmatrix}.
\end{equation}

In this basis, the electroweak scalar doublets are expressed as
\begin{equation}
\phi_{\text{vev}} =
\begin{pmatrix}
G^\pm \\
\frac{1}{\sqrt{2}} (v + \tilde{H} + i G_0)
\end{pmatrix}, \quad
\psi_1 =
\begin{pmatrix}
H_1^\pm \\
\frac{1}{\sqrt{2}} (\tilde{H}_a + i A_1)
\end{pmatrix}, \quad
\psi_2 =
\begin{pmatrix}
H_2^\pm \\
\frac{1}{\sqrt{2}} (\tilde{H}_b + i A_2)
\end{pmatrix}.
\end{equation}

While the charged and pseudoscalar sectors are already diagonal in this basis, the neutral scalar sector requires an additional rotation. The CP-even neutral scalar fields $\widetilde{H}$ and $\widetilde{H}_b$ are not mass eigenstates in general. The angle $\alpha$ is the mixing angle of the CP-even scalar sector and is responsible for diagonalizing the corresponding CP-even scalar mass matrix.  The physical mass eigenstates are obtained via

\begin{equation}
\begin{pmatrix}
\tilde{H} \\
\tilde{H}_a \\
\tilde{H}_b
\end{pmatrix}
=
\begin{pmatrix}
\cos(\alpha - \theta) & 0 & \sin(\alpha - \theta) \\
0 & 1 & 0 \\
-\sin(\alpha - \theta) & 0 & \cos(\alpha - \theta)
\end{pmatrix}
\begin{pmatrix}
H \\
h_0 \\
h
\end{pmatrix}.\label{eq1}
\end{equation} 
 Eq.~(\ref{eq1} ) shows two possible alignment scenarios:
\begin{itemize}
\item $\tilde{H} = h$: the state $h$ behaves as the SM-like Higgs boson,
\item $\tilde{H} = H$: the state  $H$  behaves as the  SM-like Higgs boson.
\end{itemize}
In our calculation, we consider $\tilde{H} = h$ as the SM-like Higgs boson, $ H $ and $ h_{0} $ represent additional scalars. In addition, a residual $Z_2$ symmetry classifies the scalar states into even and odd sectors, with $h_0$ being odd, while $\tilde{H}$ and $\tilde{H}_b$ are even as given in Table~\ref{tab:Z2_parity}. 
In our case, the alignment limit condition is given by 
\begin{equation}
\sin(\alpha - \theta) = 1, \quad \cos(\alpha - \theta) = 0.
\label{eqq}
\end{equation}

The exact alignment limit defined in Eq.(\ref{eqq}) is shown only for reference and is not imposed in the numerical calculations performed in this work. Instead, the parameters $\alpha$ and $\theta$ are considered as independent input parameters, as shown in Eq.(\ref{eq14}), and are scanned over the parameter space that satisfies all applied theoretical and experimental limits. Consequently, the compatibility of the SM-like Higgs boson with the observed Higgs signal strengths is investigated separately using HiggsSignals. Therefore, these constraints provide an additional and independent restriction on the parameter space beyond the alignment-limit consideration.

\begin{table}
\centering
\begin{tabular}{|c c|c c|c c|}
\hline
\multicolumn{2}{|c|}{Neutral scalars} & 
\multicolumn{2}{c|}{Pseudoscalars} & 
\multicolumn{2}{c|}{Charged scalars} \\
\hline
$h_0$ & odd      & $A_1$ & odd  & $H_1^\pm$ & odd \\
$\tilde{H}$ & even  & $A_2$ & even & $H_2^\pm$ & even \\
$\tilde{H}_b$ & even &        &      &            &      \\
\hline
\end{tabular}
\caption{ Assignment of  $Z_2$ parity to the physical mass eigenstates.}
\label{tab:Z2_parity}
\end{table}

We list the free parameters of the S3-3H model, namely the masses, VEVs, and  angles, used as inputs in our numerical calculations, as follows:

\begin{equation}
v, ~\, m_{h_0}, ~\, m_{H},~\, m_{h},~\, m_{A_1},~\, m_{A_2},~\, m_{H_1^\pm},~\, m_{H_2^\pm},~\, \tan\alpha,\, ~\tan\theta.
\label{eq14}
\end{equation}

\subsection{Kinetic Lagrangian and Gauge Interactions}

The gauge interactions are derived from the scalar kinetic Lagrangian \cite{GomezBock:2021},
\begin{equation}
\mathcal{L}_{\mathrm{kin}} = \sum_{i=1,2,S} (D_\mu H_i)^\dagger (D^\mu H_i),
\end{equation}
with covariant derivative 
\begin{equation}
D_\mu = \partial_\mu + i \frac{g}{2} \tau^a W_\mu^a + i \frac{g'}{2} B_\mu.
\end{equation}

After electroweak symmetry breaking, these interactions generate  the couplings of the CP-even Higgs bosons to gauge bosons.
We adopt the relevant scalar–gauge and scalar self-couplings for the phenomenology discussed in Sec.~\ref{pheno} from Ref.~\cite{GomezBock:2021}. The relevant couplings are listed in Table~\ref{tabI}. The residual  $ Z_{2} $ symmetry forbids certain couplings, and therefore they are absent.

\begin{table}[H]
\centering
\begin{tabular}{|c|c|}
\hline
\textbf{Vertex} & \textbf{Vertex factor} \\
\hline

$h_0 W^+ W^-$ & $0$ \\
\hline

$H W^+ W^-$ & $\dfrac{2 M_W^2}{v} \cos(\alpha - \theta)\, g^{\mu\nu}$ \\
\hline

$h W^+ W^-$ & $\dfrac{2 M_W^2}{v} \sin(\alpha - \theta)\, g^{\mu\nu}$ \\
\hline

$h h W^+ W^-$ & $\dfrac{M_W^2}{v^2}\, g^{\mu\nu}$ \\
\hline

$h h Z Z$ & $\dfrac{M_Z^2}{2 v^2}\, g^{\mu\nu}$ \\
\hline

$Z h A_2$ & $\dfrac{g}{2 \cos\theta_W} \cos(\alpha - \theta)\,(p + p')^\mu$ \\
\hline

$h W^\pm H_2^\mp$ & $\mp \dfrac{i g}{2} \cos(\alpha - \theta)\,(p + p')^\mu$ \\
\hline

$h h h$ & 
$-\dfrac{1}{v s_{2\theta}} \left[
m_{h_0}^2 \dfrac{c_{\alpha-\theta}^3}{9 c_\theta^2}
+ m_{h}^2 \left(c_{\alpha}^{2}c_{\alpha-\theta}- s_\alpha s_\theta \right)
\right]$ \\
\hline

$H h h$ & 
$\dfrac{c_{\alpha-\theta}}{v s_{2\theta}} \left[
m_{h_0}^2 \left(\dfrac{s_{2(\alpha-\theta)}}{6 c_\theta^2}\right)
+ \dfrac{m_{H}^2 s_{2\alpha}}{2}
+ m_{h}^2 s_{2\alpha}
\right]$ \\
\hline
\end{tabular}
\caption{Relevant scalar--gauge and scalar self-interaction couplings of the S3-3H model in the physical basis.}
\label{tabI}
\end{table}
where $c_{\alpha-\theta} = \cos(\alpha - \theta)$ and $s_{2\theta} = \sin(2\theta)$. Note that the gauge boson couplings to neutral CP-even scalars ($ h,H $) in the S3-3H model are similar to those in the  Two Higgs Doublet Model (2HDM).

\section{CONSTRAINTS}

The S3-3H model is restricted by theoretical considerations such as perturbative unitarity and vacuum stability, as well as by experimental limits obtained from collider experiments such as the Tevatron and the LHC.  We first investigate constraints on model parameters from theoretical considerations and then those from the collider experiments.

\subsection{Constraints from Unitarity and Stability}
\begin{figure}[t]
\begin{center}
\includegraphics[width=0.23\linewidth]{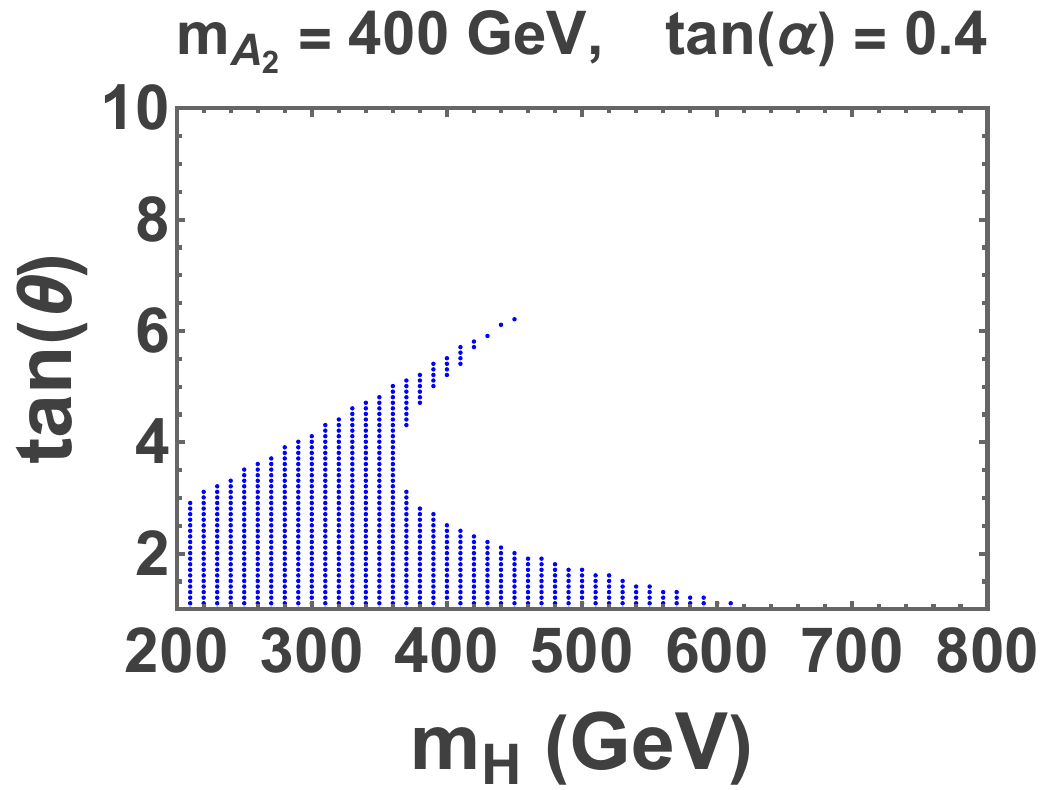}\hspace{3.25mm}
\includegraphics[width=0.23\linewidth]{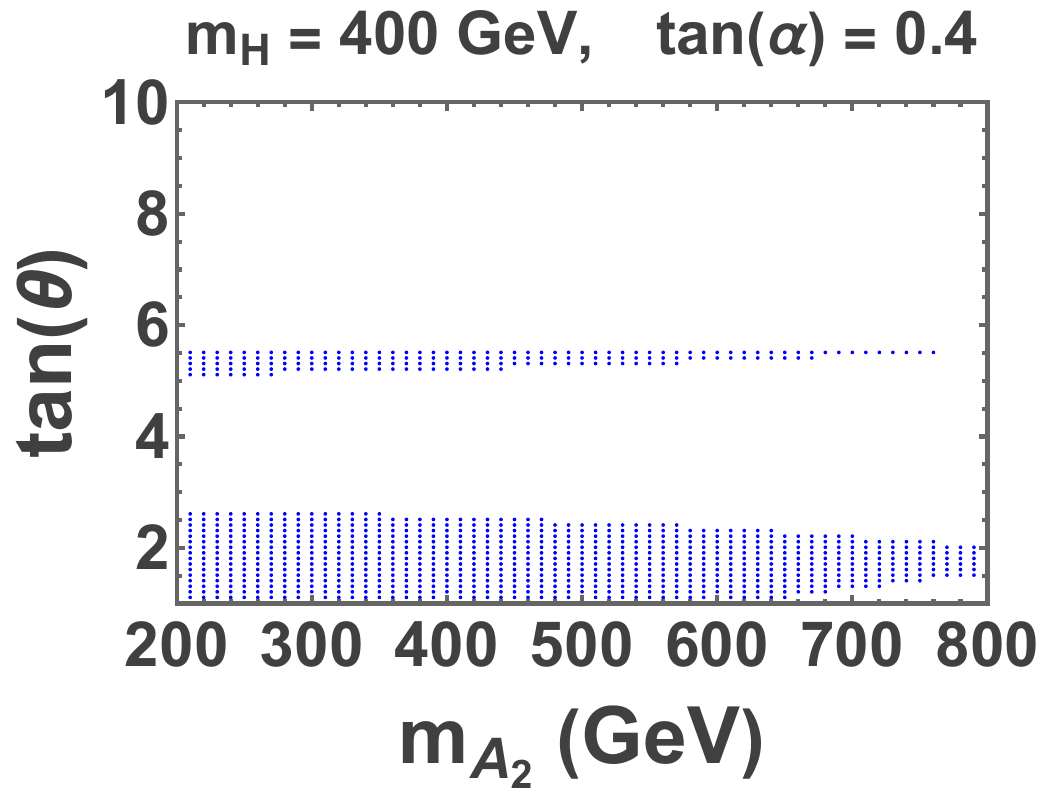}\hspace{3.25mm}
\includegraphics[width=0.23\linewidth]{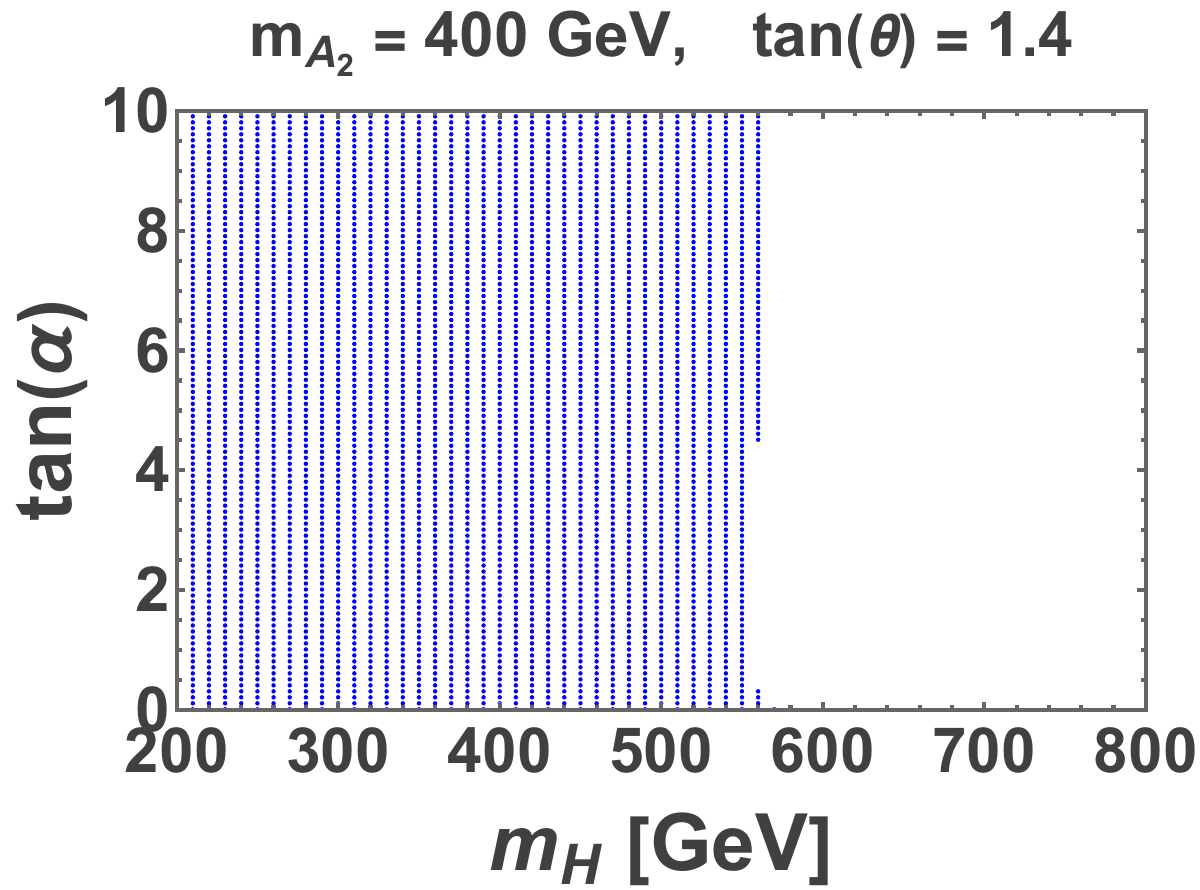}\hspace{3.25mm}
\includegraphics[width=0.225\linewidth]{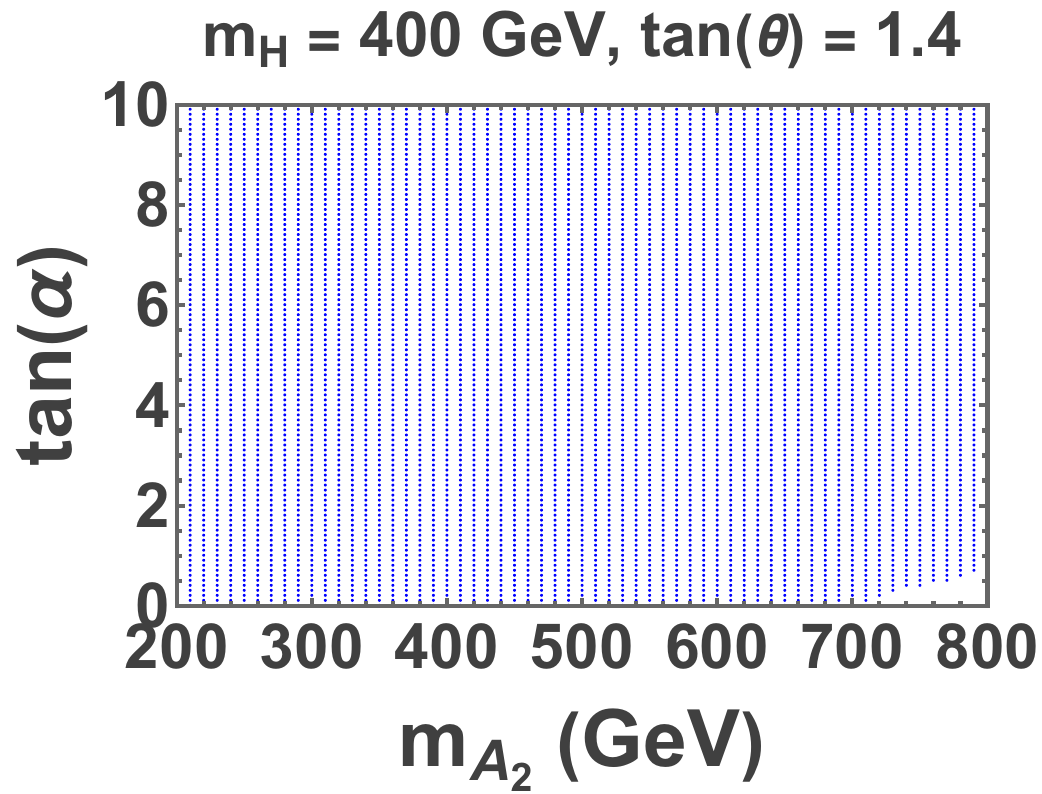}\hspace{3.25mm}

\caption{Blue: Allowed  regions from unitarity and stability  in the  S3-3H model. As for the reference input values, we take $m_h = 125~\text{GeV},\; m_{h_0} = 200~\text{GeV},~ m_{H_1^\pm}= m_{H_2^\pm}= 350~\text{GeV}$  and $ m_{A_1} = 400~\text{GeV}.$ }
\label{fig1}
\end{center}
\end{figure}

The parameter space of the model is  constrained by theoretical requirements such as perturbative unitarity and vacuum stability. We adopt the  definitions for perturbative unitarity bounds from  Ref.~\cite{66}, since our potential structure is similar to  that of Case II discussed therein. Vacuum stability is imposed by requiring the scalar potential to be bounded from below, which leads to positivity conditions on the quartic couplings. In Fig.~\ref{fig1}, the parameter space on the $(m_H,\tan\theta)$, $(m_{A_2},\tan\theta)$,  $(m_H,\tan\alpha)$, and $(m_{A_2},\tan\alpha)$    planes, shown by blue points, respects perturbative unitarity and vacuum stability. 

\subsection{Constraints from Collider Experiments}

\begin{figure}[t]
\begin{center}
\includegraphics[width=0.48\linewidth]{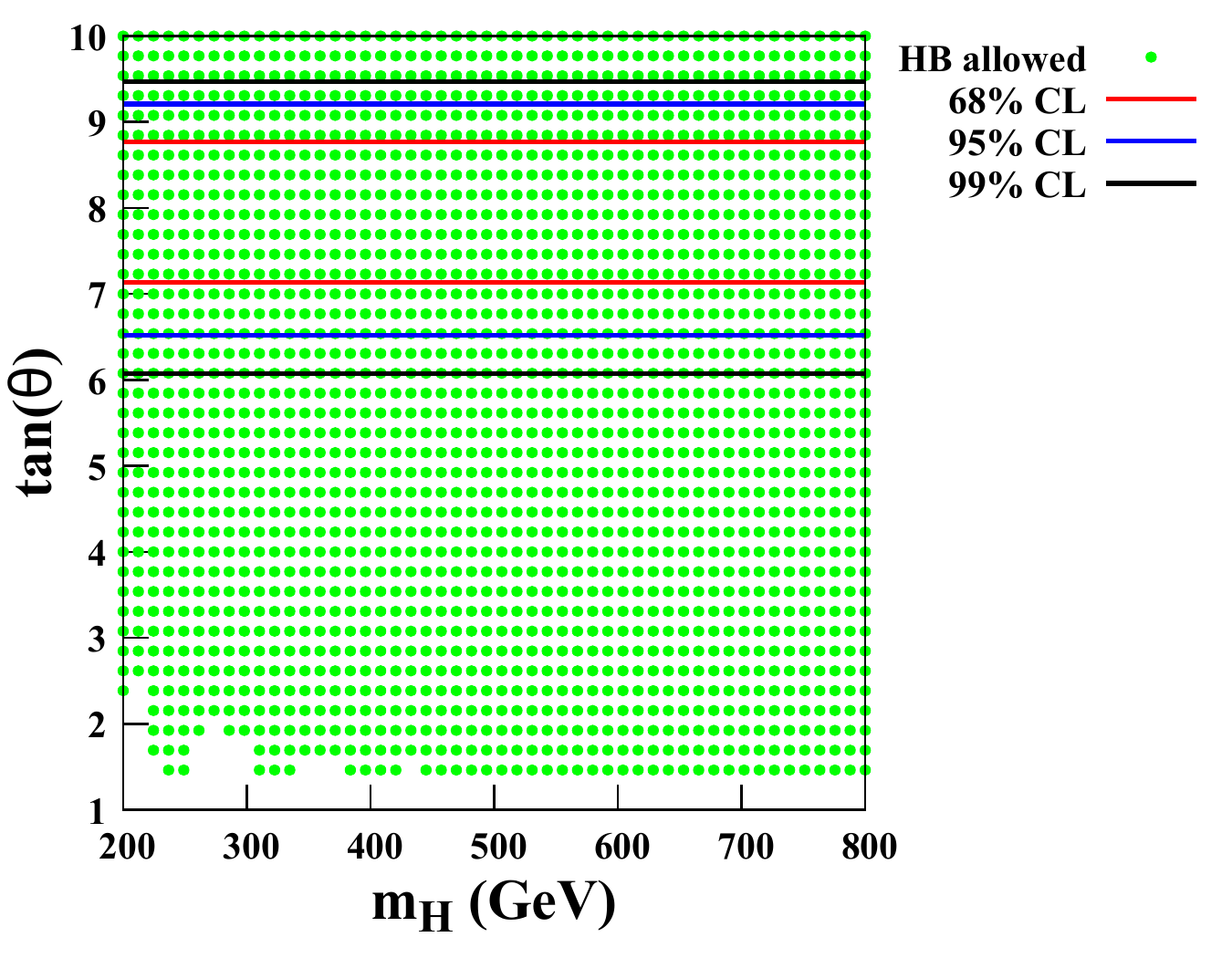}\hspace{3.25mm}
\includegraphics[width=0.48\linewidth]{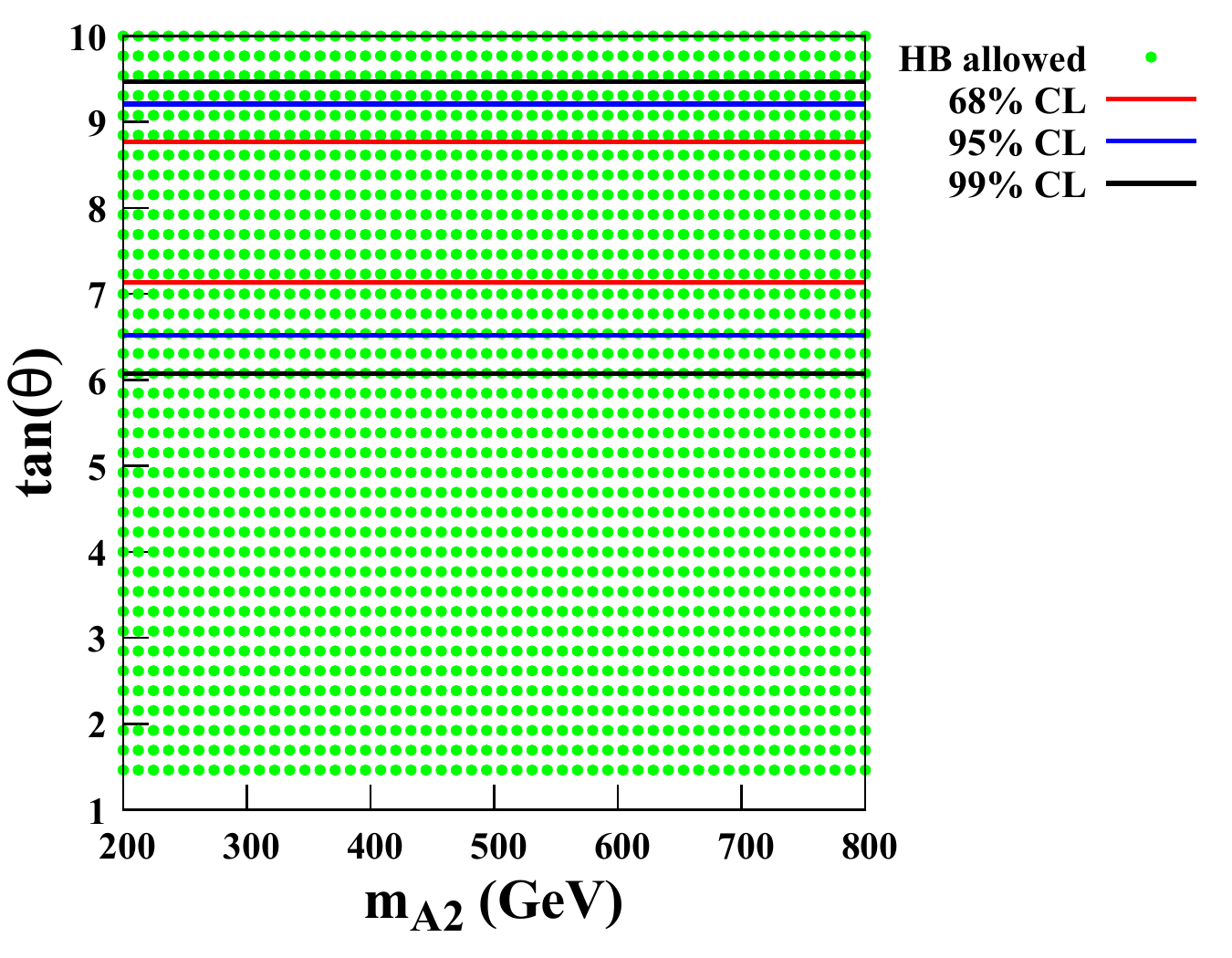}\hspace{3.25mm}

\caption{Green: Allowed regions from collider experiments. The red, blue and black   contours illustrate  the compatibility with observed
Higgs signals (SM signal strengths) at $\Delta \chi^2 = 68\%,\,95\%,\,99\%~\mathrm{CL} $ respectively.  We take the reference input values: $m_{h}=125$ GeV,  $m_{h_0}=200$ GeV,   $ m_{H_{2}^{\pm}}=350$ GeV,  $ m_{A_{1}}=m_{A_2}=m_{H}=400 $ GeV and~  $\tan \alpha=0.4$. }
\label{fig2}
\end{center}
\end{figure}

\begin{figure}[t]
\begin{center}

\includegraphics[width=0.48\linewidth]{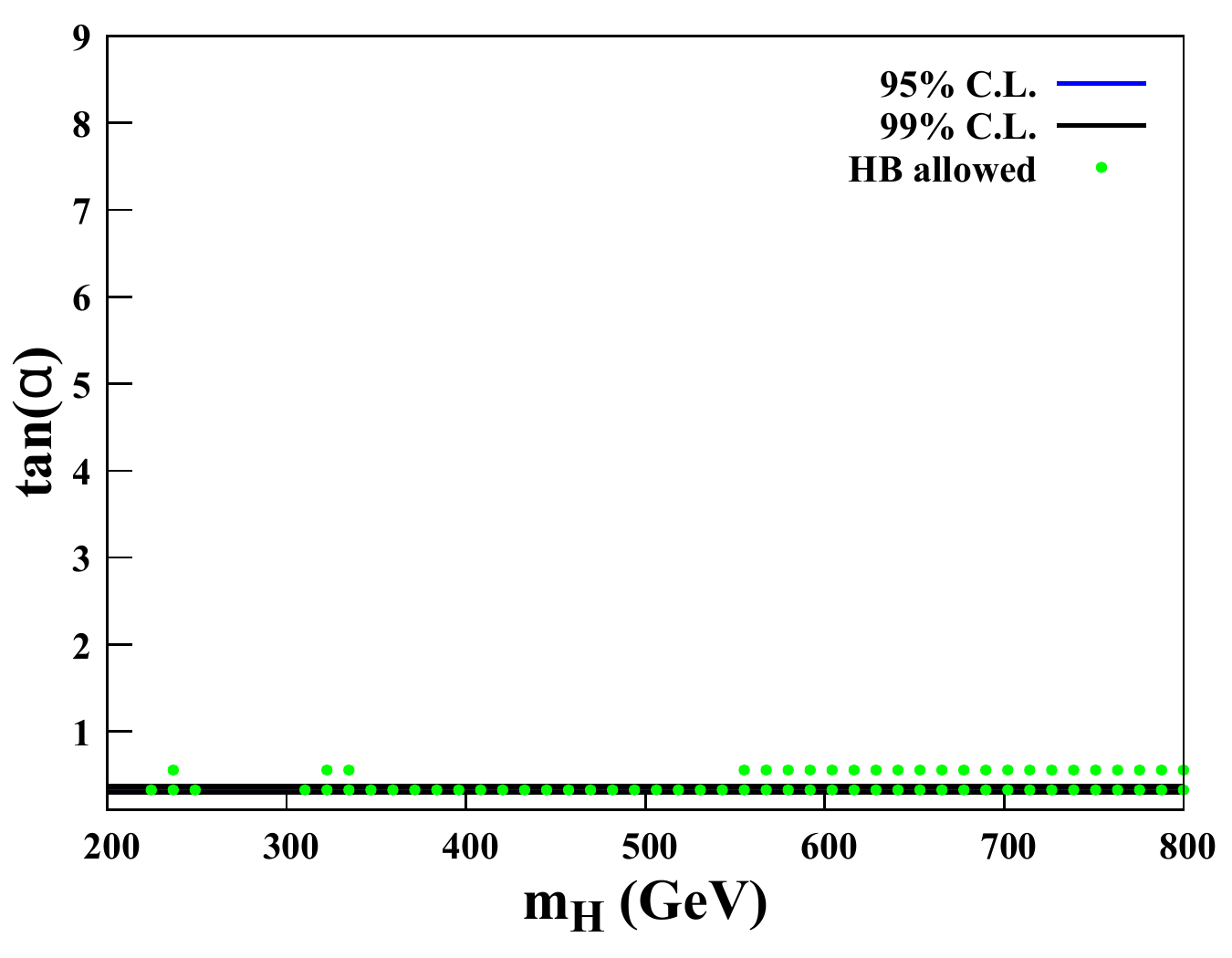}\hspace{3.25mm}
\includegraphics[width=0.48\linewidth]{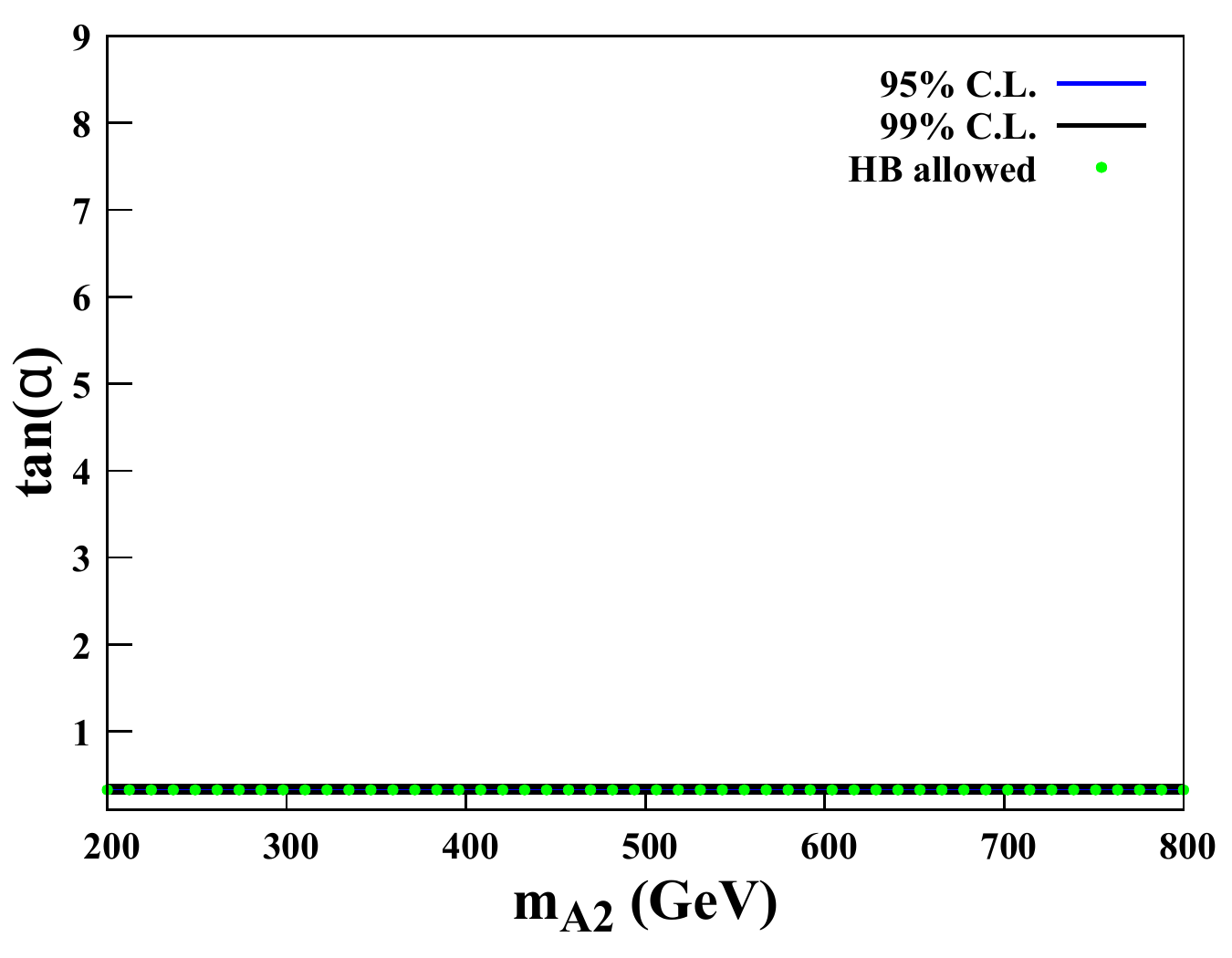}\hspace{3.25mm}

\caption{Green: Allowed regions from collider experiments. The  blue and black   contours illustrate  the compatibility with observed Higgs signals (SM signal strengths) at $\Delta \chi^2 = 95\%~ \text{and}~ \,99\%~\mathrm{CL} $ respectively.  We take the reference input values: $m_{h}=125$ GeV,  $m_{h_0}=200$ GeV,   $ m_{H_{2}^{\pm}}=350$ GeV,  $ m_{A_{1}}=m_{A_2}=m_{H}=400 $ GeV and~  $\tan \theta=1.4$. The allowed parameter space is confined to a narrow interval around $\tan\alpha \simeq 0.4. $ }
\label{fig2-add}
\end{center}
\end{figure}

In this subsection, we take into account constraints on  parameters such as $ m_{H}$, $ m_{A_{2} }$, tan($\theta$) tan$ (\alpha )$ of the S3-3H model from data, collected at the  Tevatron and  the LHC by using the {\tt Higgs Bounds} (v5.3.2 beta) package~\cite{HB1,HB2,HB3,HB4}. We also investigate the compatibility of the signal strengths of the SM-like Higgs boson through a  $\Delta \chi^2$ analysis employing the HiggsSignals \cite{HS} (v2.2.3 beta) package. In Figs.~\ref{fig2} and \ref{fig2-add}, the green region represents the parameter space allowed by collider experiments. The red, blue, and black contours correspond to the $68.27\%$, $95.45\%$, and $99.73\%$ confidence level regions, respectively, obtained from the $\Delta \chi^2$ analysis in the $(m_H,\tan\theta)$, $(m_{A_2},\tan\theta)$,  $(m_H,\tan\alpha)$, and $(m_{A_2},\tan\alpha)$  planes.

The parameter space of the S3-3H model under various theoretical and experimental constraints, including collider limits, has been studied in the literature~ \cite{Espinoza:2018itz, GomezBock:2021, 66}. The aim of our study is not to perform a new parameter-space analysis. Instead, we use the experimentally and theoretically allowed regions to study double Higgs  production at future $e^+e^-$ colliders. In particular, we investigate the processes $e^+e^- \to hhZ$ and $e^+e^- \to hh\nu_e\bar{\nu}_e$, determine the impact of the additional scalar states on these channels, and discuss the mechanisms responsible for the observed cross-section enhancements.

The parameter space was explored through dedicated scans in the
$(m_H,\tan\theta)$, $(m_{A_2},\tan\theta)$, $(m_H,\tan\alpha)$, and $(m_{A_2},\tan\alpha)$ planes. For each scan, approximately 6000 input points were tested against the perturbative unitarity and vacuum stability constraints, while 2000 points were tested using HiggsBounds. The surviving points   were obtained from the intersection of the points satisfying the theoretical constraints and those allowed by HiggsBounds. After applying all theoretical and experimental constraints, only about $2.0$--$2.5\%$ of the scanned points remained allowed, depending on the parameter plane considered. We illustrate these surviving solutions with two benchmark scenarios given in Table~\ref{tab:benchmark}.

\begin{table}[ht]
\centering
\renewcommand{\arraystretch}{1.2}
\begin{tabular}{|c|c|c|}
\hline
Benchmark Scenario &
Fixed Parameters &
Allowed Parameter Region \\
\hline

Benchmark Scenario I &
$m_H = 400~\mathrm{GeV}$ &
$200 < m_{A_2} < 700~\mathrm{GeV}$, \;
$\tan\theta=1.4$, \; $\tan\alpha=0.4$ \\
\hline

Benchmark Scenario II &
$m_{A_2} = 400~\mathrm{GeV}$ &
$310 < m_H < 560~\mathrm{GeV}$, \;
$\tan\theta = 1.4$,\; $\tan\alpha=0.4$ \\
\hline
\end{tabular}

\caption{Representative benchmark scenarios selected from the largest continuous allowed regions of parameter space after imposing perturbative unitarity, vacuum stability, HiggsBounds constraints, taking into account the HiggsSignals compatibility contours  shown in Figs.~\ref{fig2} and \ref{fig2-add}.}
\label{tab:benchmark}
\end{table}

Next, we will calculate the production cross sections at $ e^{+}e^{-} $ colliders by applying  the benchmark {scenarios} given in Table~\ref{tab:benchmark}.
\section{PHENOMENOLOGY} \label{pheno}

\subsection{Double Higgs Production in the S3-3H Model}
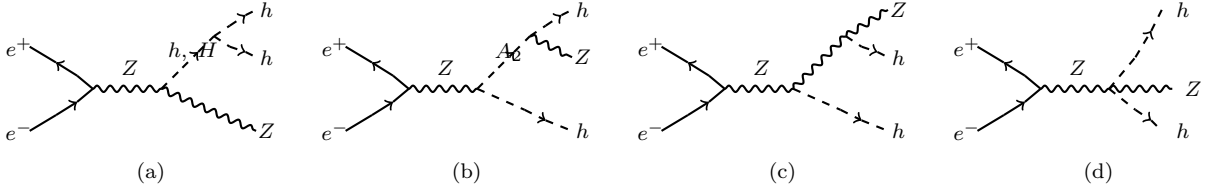
\begin{figure}[htbp]
\centering
\begin{tikzpicture}[scale=0.70, xshift=-2.5cm, every node/.style={font=\scriptsize}]
\tikzset{
  ferm/.style={thick},
  scalar/.style={thick,dashed},
  boson/.style={thick,decorate,decoration={snake,amplitude=1.2pt,segment length=5pt}}
}

\begin{scope}[xshift=0cm]

\draw[ferm] (0,0) -- (0.5,-0.3);
\draw[ferm,<-] (0.5,-0.3) -- (0.9,-0.55);
\draw[ferm] (0.9,-0.55) -- (1.2,-0.8);

\draw[ferm] (0,-1.6) -- (0.5,-1.3);
\draw[ferm,->] (0.5,-1.3) -- (0.9,-1.05);
\draw[ferm] (0.9,-1.05) -- (1.2,-0.8);

\node at (-0.2,0) {$e^+$};
\node at (-0.2,-1.6) {$e^-$};

\draw[boson] (1.2,-0.8) -- (2.5,-0.8);
\node at (1.9,-0.4) {$Z$};

\draw[scalar] (2.5,-0.8) -- (2.9,-0.4);
\draw[scalar,->] (2.9,-0.4) -- (3.2,-0.1);
\draw[scalar] (3.2,-0.1) -- (3.5,0.2);

\node[anchor=east] at (4.04,-0.55) {$h,\;H$};

\draw[scalar] (3.5,0.2) -- (3.8,0.4);
\draw[scalar,->] (3.8,0.4) -- (4.1,0.6);
\draw[scalar] (4.1,0.6) -- (4.3,0.7);

\draw[scalar] (3.5,0.2) -- (3.8,0.0);
\draw[scalar,->] (3.8,0.0) -- (4.1,-0.1);
\draw[scalar] (4.1,-0.1) -- (4.3,-0.2);

\draw[boson] (2.5,-0.8) -- (4.3,-1.6);

\node at (4.5,0.7) {$h$};
\node at (4.5,-0.2) {$h$};
\node at (4.5,-1.6) {$Z$};
\node at (2.3,-2.4) {(a)};

\end{scope}

\begin{scope}[xshift=6cm]

\draw[ferm] (0,0) -- (0.5,-0.3);
\draw[ferm,<-] (0.5,-0.3) -- (0.9,-0.55);
\draw[ferm] (0.9,-0.55) -- (1.2,-0.8);

\draw[ferm] (0,-1.6) -- (0.5,-1.3);
\draw[ferm,->] (0.5,-1.3) -- (0.9,-1.05);
\draw[ferm] (0.9,-1.05) -- (1.2,-0.8);

\node at (-0.2,0) {$e^+$};
\node at (-0.2,-1.6) {$e^-$};

\draw[boson] (1.2,-0.8) -- (2.5,-0.8);
\node at (1.9,-0.4) {$Z$};

\draw[scalar] (2.5,-0.8) -- (2.9,-0.4);
\draw[scalar,->] (2.9,-0.4) -- (3.2,-0.1);
\draw[scalar] (3.2,-0.1) -- (3.5,0.2);

\node[anchor=east] at (3.8,-0.37) {$A_2$};

\draw[scalar] (3.5,0.2) -- (3.8,0.4);
\draw[scalar,->] (3.8,0.4) -- (4.1,0.6);
\draw[scalar] (4.1,0.6) -- (4.3,0.7);

\draw[boson] (3.5,0.2) -- (4.3,-0.2);

\draw[scalar] (2.5,-0.8) -- (3.4,-1.2);
\draw[scalar,->] (3.4,-1.2) -- (3.8,-1.4);
\draw[scalar] (3.8,-1.4) -- (4.3,-1.6);

\node at (4.5,0.7) {$h$};
\node at (4.5,-0.2) {$Z$};
\node at (4.5,-1.6) {$h$};
\node at (2.3,-2.4) {(b)};

\end{scope}
\begin{scope}[xshift=12cm]

\draw[ferm] (0,0) -- (0.5,-0.3);
\draw[ferm,<-] (0.5,-0.3) -- (0.9,-0.55);
\draw[ferm] (0.9,-0.55) -- (1.2,-0.8);

\draw[ferm] (0,-1.6) -- (0.5,-1.3);
\draw[ferm,->] (0.5,-1.3) -- (0.9,-1.05);
\draw[ferm] (0.9,-1.05) -- (1.2,-0.8);

\node at (-0.2,0) {$e^+$};
\node at (-0.2,-1.6) {$e^-$};

\draw[boson] (1.2,-0.8) -- (2.5,-0.8);
\node at (1.9,-0.4) {$Z$};

\draw[boson] (2.5,-0.8) -- (3.5,0.2);
\draw[boson] (3.5,0.2) -- (4.3,0.7);

\draw[scalar] (3.5,0.2) -- (3.8,0.0);
\draw[scalar,->] (3.8,0.0) -- (4.1,-0.1);
\draw[scalar] (4.1,-0.1) -- (4.3,-0.2);

\draw[scalar] (2.5,-0.8) -- (3.4,-1.2);
\draw[scalar,->] (3.4,-1.2) -- (3.8,-1.4);
\draw[scalar] (3.8,-1.4) -- (4.3,-1.6);

\node at (4.5,0.7) {$Z$};
\node at (4.5,-0.2) {$h$};
\node at (4.5,-1.6) {$h$};
\node at (2.3,-2.4) {(c)};
\end{scope}

\begin{scope}[xshift=18cm]

\draw[ferm] (0,0) -- (0.5,-0.3);
\draw[ferm,<-] (0.5,-0.3) -- (0.9,-0.55);
\draw[ferm] (0.9,-0.55) -- (1.2,-0.8);

\draw[ferm] (0,-1.6) -- (0.5,-1.3);
\draw[ferm,->] (0.5,-1.3) -- (0.9,-1.05);
\draw[ferm] (0.9,-1.05) -- (1.2,-0.8);

\node at (-0.2,0) {$e^+$};
\node at (-0.2,-1.6) {$e^-$};

\draw[boson] (1.2,-0.8) -- (2.5,-0.8);
\node at (1.9,-0.4) {$Z$};

\draw[scalar] (2.5,-0.8) -- (3.0,-0.2);
\draw[scalar,->] (3.0,-0.2) -- (3.3,0.3);
\draw[scalar] (3.3,0.3) -- (3.5,0.7);

\draw[scalar] (2.5,-0.8) -- (3.0,-1.2);
\draw[scalar,->] (3.0,-1.2) -- (3.3,-1.4);
\draw[scalar] (3.3,-1.4) -- (3.5,-1.6);

\draw[boson] (2.5,-0.8) -- (3.7,-0.8);

\node at (3.9,0.7) {$h$};
\node at (3.9,-1.6) {$h$};
\node at (4.1,-0.8) {$Z$};
\node at (2.3,-2.4) {(d)};
\end{scope}

\end{tikzpicture}
\caption{Feynman diagrams for $e^+e^- \to hhZ$ production.}
\label{fig:hhZ}
\end{figure}

In this subsection, we present the numerical results for the cross sections of double Higgs production via the Higgs-strahlung (HS) process $(e^+ e^- \rightarrow hhZ) $ in Fig.~\ref{fig:hhZ} and the Vector Boson Fusion (VBF) mechanism  $ (e^+ e^- \rightarrow  h h\nu_{e} \bar{\nu}_{e})$ in Fig.~\ref{fig:vvhh_feynman}.

\subsubsection{Double Higgs-Strahlung }
The differential cross section for the process 
$(e^+ e^- \rightarrow hhZ )$ is taken from Refs.~\cite{djouadi,djouadi2,osland, Li:2017daq} and can be defined as

\begin{equation}
\frac{d\sigma(e^+ e^- \rightarrow hhZ )}{dx_1 dx_2} =
\frac{G_F^3 m_Z^6}{384\sqrt{2}\pi^3 s}        
\left( V_e^2 + A_e^2 \right)
\frac{\mathcal{A}}{(1 - r_Z)^2}, \label{eqq3}
\end{equation}

where $x_{1,2} = 2E_{1,2}/\sqrt{s}$ stand for the energies of the final-state Higgs boson and  $s$ is the center-of-mass energy squared. $x_3 = 2 - x_1 - x_2$ and $y_i = 1 - x_i$ with $i = 1,2,3$ are shorthand notations. Furthermore, the mass ratios are introduced as $r_{i} = m_{i}^2/s$. Ref.~\cite{osland} defines the function $\mathcal{A}$ that  includes the contributions from the different Feynman diagrams, including the trilinear Higgs self-coupling and gauge interactions as 
\begin{equation}
\mathcal{A} = r_Z \left\{
\frac{1}{2} |a|^2 f_a
+ |b(y_1)|^2 f_b
+ 2\,\mathrm{Re}\left[a\, b^*(y_1)\right] g_{ab}
+ \mathrm{Re}\left[b(y_1)\, b^*(y_2)\right] g_{bb}
\right\}
+ \{ x_1, y_1 \leftrightarrow x_2, y_2 \}. \label{eq3}
\end{equation}
where
\begin{equation}
a = \frac{1}{2} \left[
\left(
\frac{\kappa_{hVV}\,\kappa_{hhh}}{y_3 + r_Z - \tilde{r}_{h}}
+
\frac{\kappa_{H VV}\,\kappa_{H hh}}{y_3 + r_Z - \tilde{r}_{H}}
\right)
\frac{3 m_{h}^2}{m_Z^2}
+
\left(
\frac{\kappa_{hVV}^2}{y_1 + r_h - \tilde{r}_Z}
+
\frac{\kappa_{h VV}^2}{y_2 + r_h - \tilde{r}_Z}
\right)
+
\frac{\kappa_{h hVV}}{2 r_Z}
\right],
\end{equation}

\begin{equation}
b(y) = \frac{1}{2 r_Z}
\left(
\frac{\kappa_{h VV}^2}{y + r_h - \tilde{r}_Z}
+
\frac{\kappa_{A_2 h Z}^2}{y + r_h - \tilde{r}_{A_2}}
\right).
\end{equation}

The propagator factors are expressed as
\begin{equation}
\tilde{r}_i = \frac{m_i^2 - i m_i \Gamma_i}{s}, \quad i = h, Z, H, A_2.
\end{equation}

The explicit expressions of  the coefficients $f_a$, $f_b$, $g_{ab}$, and $g_{bb}$ appearing in Eq.~(\ref{eq3}) are given in Appendix~\ref{app:A}. Here, the coupling modifiers are defined as $(\kappa_{XYZ} \equiv g^{\mathrm{S3\text{-}3H}}_{XYZ}/g^{\mathrm{SM}}_{XYZ})$  which represent the ratio of the corresponding coupling in the S3-3H model to its  SM value.

\begin{figure}[H]
\begin{center}
\includegraphics[width=0.35\linewidth]{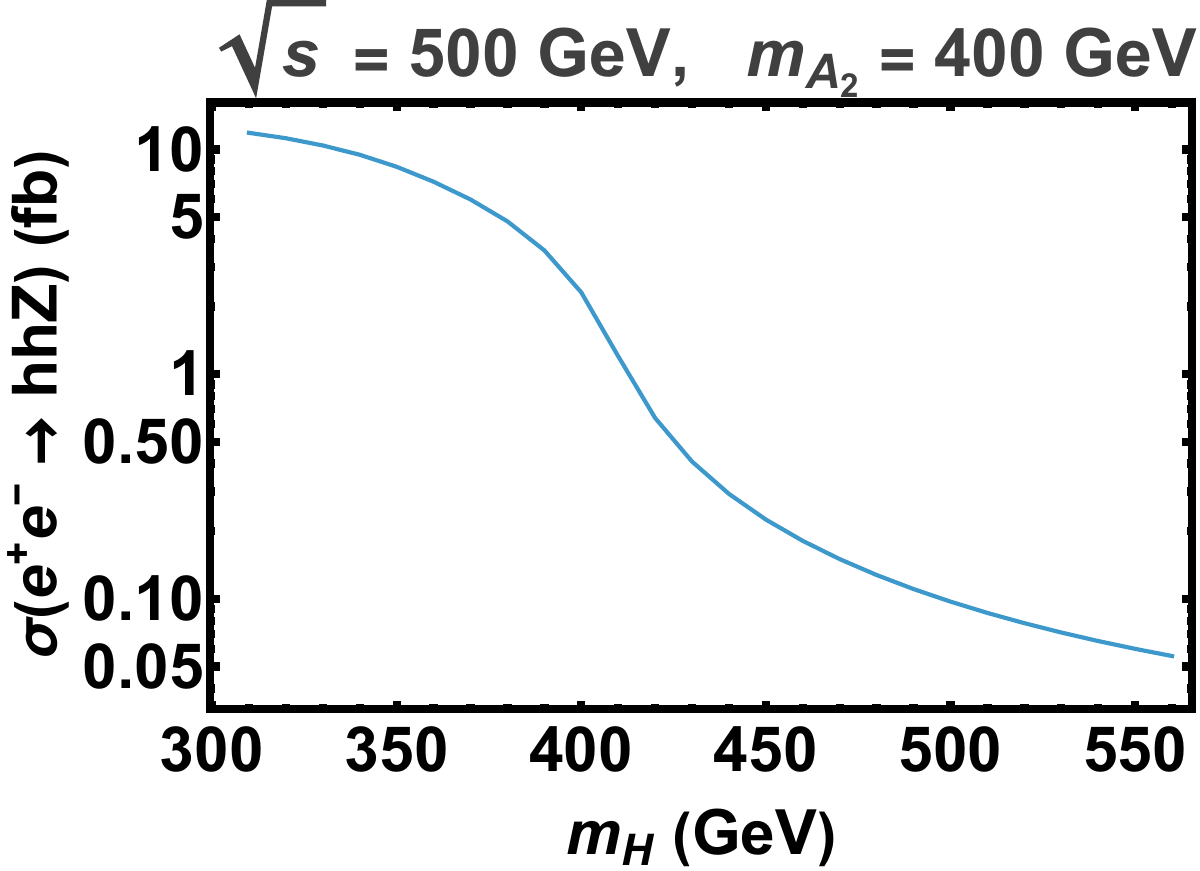}\hspace{3.25mm}
\includegraphics[width=0.35\linewidth]{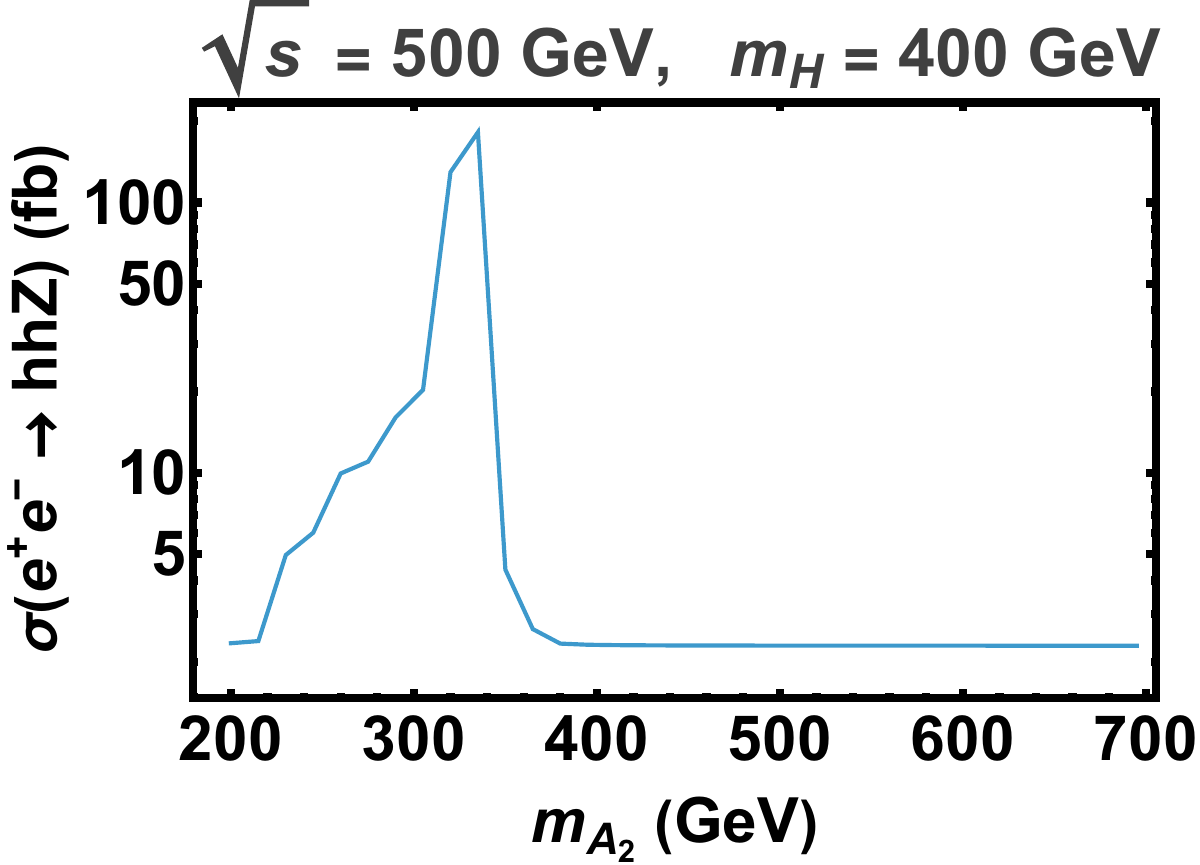}\hspace{3.25mm}\\    \vspace{5mm}
\includegraphics[width=0.35\linewidth]{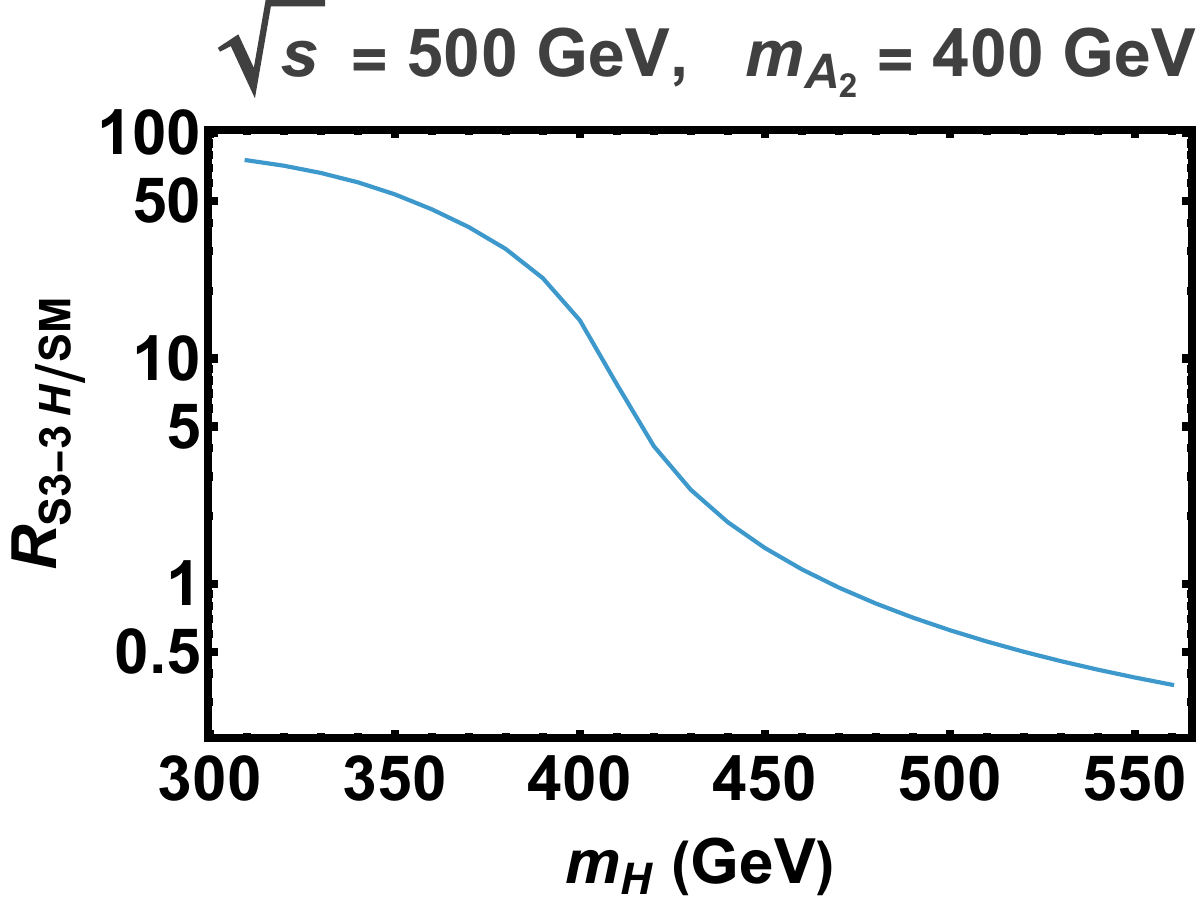}\hspace{3.25mm}
\includegraphics[width=0.36\linewidth]{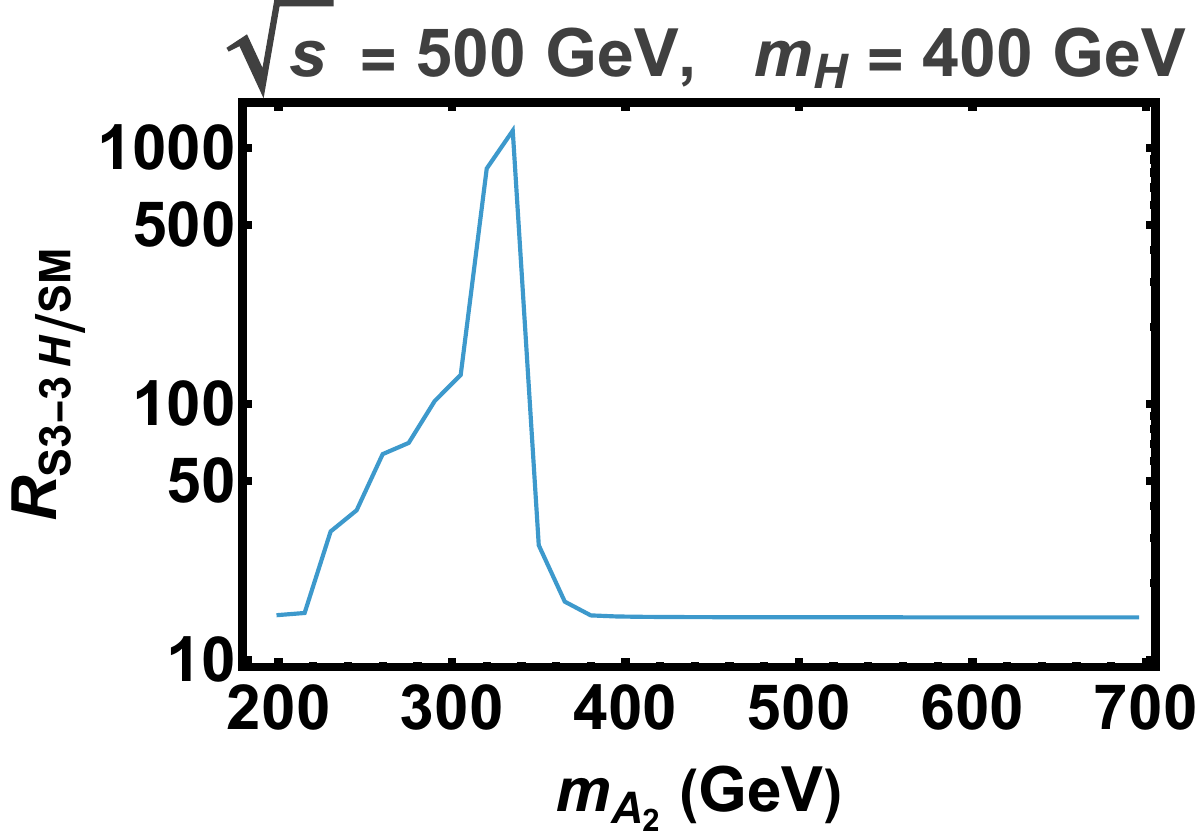}\hspace{3.25mm}\\    \vspace{5mm}

\caption{The upper panels: Cross section for the process $e^+e^- \to hhZ$  as a function of $m_{H}$ (left) and $m_{A_2}$ (right),  Lower panels: The  ratio $R=\sigma_{\text{S3-3H}}/\sigma_{\text{SM}}$.    We take $m_{h}=125$ GeV,  $m_{h_0}=200$ GeV, $\tan\alpha=0.4$ and $\tan\theta=1.4$. }
\label{fig22}
\end{center}
\end{figure}

We obtain the  total cross section  by integrating  Eq.~(\ref{eqq3})  over the allowed kinematic phase space:
\begin{equation}
\sigma = \int dx_1 dx_2 \; \frac{d\sigma}{dx_1 dx_2}.
\end{equation}
Next, we numerically evaluate the total cross section for  $ e^+e^- \rightarrow hhZ $, emphasizing the effect of additional scalar contributions and their interplay with the trilinear Higgs self-coupling in Figs.~\ref{fig22} and \ref{fig3}.

The  left panel of Fig.~\ref{fig22} shows the dependence of the cross section on the heavy Higgs mass $m_{H}$. 
The cross section decreases with increasing $m_{H}$ due to the suppression of the available phase space and propagator suppression effects at higher masses. In the left panel, the resonant contribution of the heavy Higgs boson is negligible. The right panel indicates the dependence of the cross section on $m_{A_2}$. A clear peak appears around $m_{A_2} \sim 300$--$350~\text{GeV}$. This enhancement arises from the intermediate CP-odd scalar $A_2$, which contributes through the $A_2 h Z$ coupling. In the relevant kinematic region, this contribution becomes significant and causes an increase in the $e^+e^- \to hhZ$ cross section. These results show that  the enhancement is mainly driven by contributions from additional  Higgs states, while the non-enhanced region is dominated by kinematic suppression effects.

The lower panels in Fig.~\ref{fig22} show the ratio $R = \sigma^{\mathrm{S3-3H}} / \sigma^{\mathrm{SM}}$ as a function of the scalar masses. This ratio exceeds unity by several orders of magnitude in certain regions of the parameter space. These results show that the enhancement in the cross section is mainly driven by contributions from additional scalar states when they become kinematically accessible, rather than by modifications of the Higgs self-coupling.  A strong localized  enhancement is  observed in the $m_{A_2}$ scan, where the contribution of $A_2$ gives rise to very large values of $R$. A clear peak appears around $m_{A_2}\sim 300$--$350$ GeV. Since the SM-like Higgs boson mass is fixed to $m_h=125$ GeV, the process $e^+e^- \rightarrow hA_2$, followed by the decay $A_2 \rightarrow Zh$, becomes kinematically allowed for $m_{A_2}>m_Z+m_h\simeq216$ GeV. The enhancement starts above this threshold and is mainly due to the resonant contribution of the CP-odd scalar $A_2$ through the $A_2Zh$ coupling.  However, the behavior in the $m_H$ scan does not present a pronounced resonant structure, and the enhancement in this case is mainly governed by phase space and propagator suppression effects. The correction $\Delta R = (\sigma^{\mathrm{S3-3H}} - \sigma^{\mathrm{SM}})/\sigma^{\mathrm{SM}}$ reaches values of up to $\mathcal{O}(10^2)$ in the $m_H$ scan, while it increases up to $\mathcal{O}(10^3)$ in the $m_{A_2}$ scan. 

\begin{figure}[H]
\begin{center}
\includegraphics[width=0.56\linewidth]{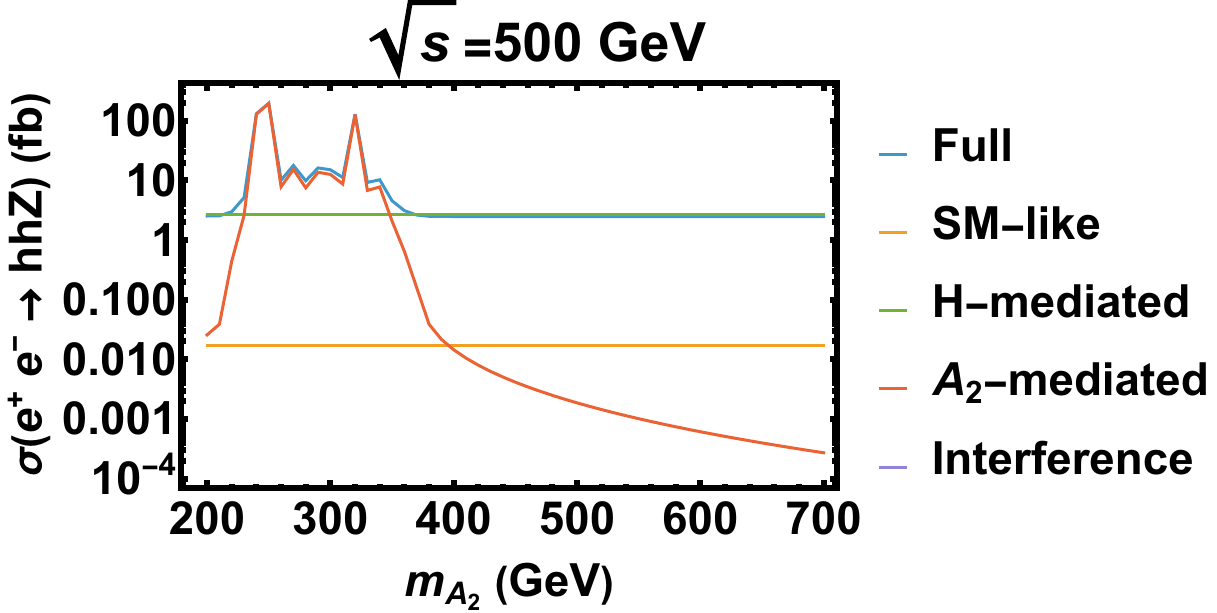}\hspace{3.25mm}
\includegraphics[width=0.37\linewidth]{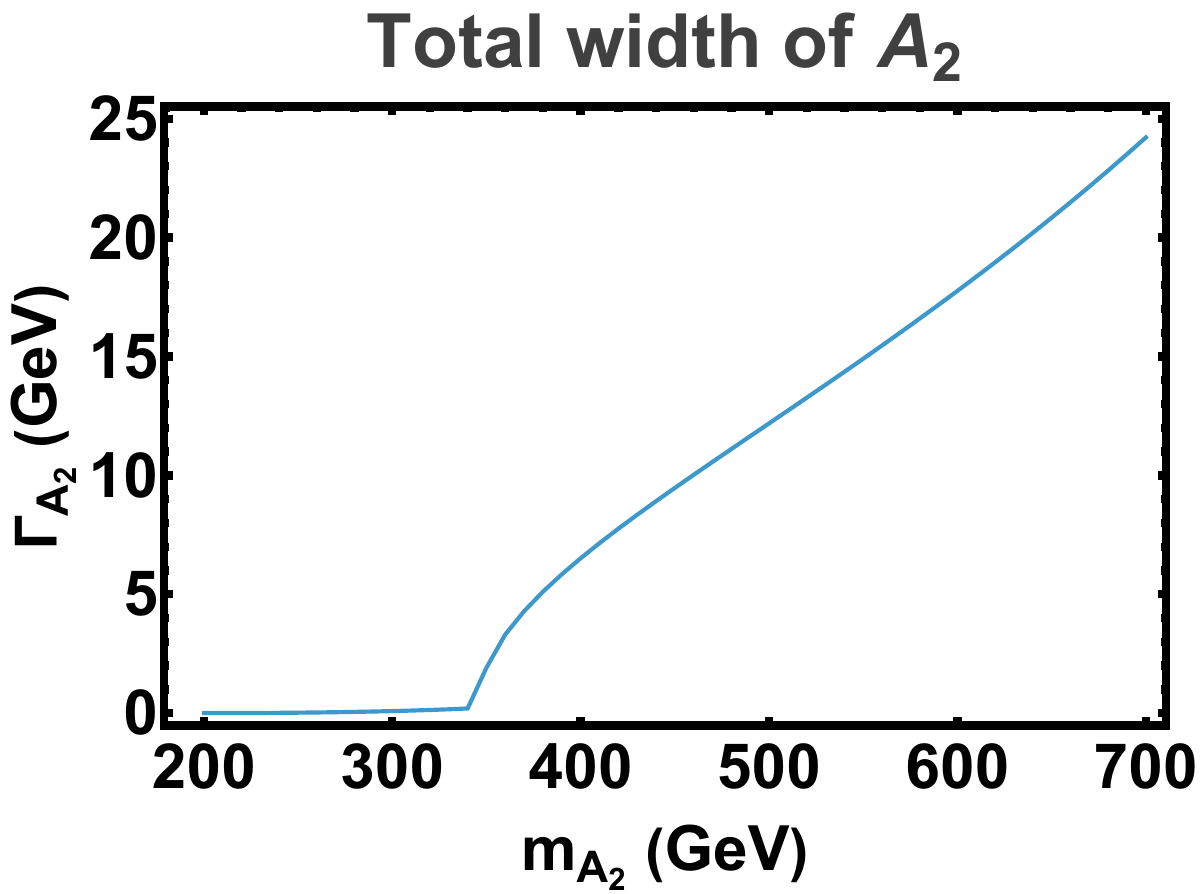}\hspace{3.5mm}\\   

\caption{Left panel: Decomposition of the $e^{+}e^{-}\to hhZ$ cross section into SM-like, $H$-mediated, $A_{2}$-mediated, and interference contributions as a function of $m_{A_2}$ at $\sqrt{s}=500$ GeV. Right panel: Total decay width $\Gamma_{A_2}$ as a function of $m_{A_2}$. The input parameters are $m_h=125$ GeV, $m_{h_0}=200$ GeV, $m_H=400$ GeV, $\tan\alpha=0.4$, and $\tan\theta=1.4$.}   
\label{decompose}
\end{center}
\end{figure}

To discuss the origin of the large enhancement observed in the right panel of Fig.~\ref{fig22}, we decompose the total cross section into SM-like, $H$-mediated, $A_2$-mediated, and interference contributions. The left panel of Fig.~\ref{decompose} shows that the enhancement is mostly generated by the $A_2$-mediated contribution. For instance, at $m_{A_2}=340$ GeV, the total cross section is $\sigma_{\rm Full}=10.27$ fb, while the individual contributions are $\sigma_{A_2}=7.78$ fb, $\sigma_H=2.72$ fb, and $\sigma_{\rm SM-like}=0.017$ fb. The interference contribution remains small and slightly destructive, $\sigma_{\rm int}=-0.25$ fb.

The right panel of Fig.~\ref{decompose} displays the total width $\Gamma_{A_2}$ as a function of $m_{A_2}$.  In this mass range, the resonant contribution of the $A_2$ propagator is maximized while the total width remains relatively small. As $m_{A_2}$ increases further, the rapid growth of $\Gamma_{A_2}$ suppresses the resonant contribution, leading to a reduction of the cross section.
 In the region where the enhancement is maximal, the pseudoscalar $A_2$ remains a narrow resonance. For example, at $m_{A_2}=260$ GeV and $340$ GeV, the corresponding widths are only $\Gamma_{A_2}=0.03$ GeV and $0.20$ GeV, respectively. In contrast, at $m_{A_2}=400$ GeV, the width increases to $\Gamma_{A_2}=6.51$ GeV, while the $A_2$-mediated contribution decreases dramatically to $\sigma_{A_2}=0.015$ fb.

These results indicate that the large values of
$R=\sigma_{\rm S3-3H}/\sigma_{\rm SM}$ originate primarily from the resonant propagation of the CP-odd scalar
$A_2$ in the narrow-width regime. The enhancement is therefore not driven by interference effects, which remain numerically small, nor by modifications of the SM-like Higgs self-coupling. As additional decay channels become kinematically accessible and $\Gamma_{A_2}$ increases, the resonant contribution is progressively suppressed, leading to a significant reduction of the cross section.

In the left  and middle panels of Fig.~\ref{fig3}, we indicate the cross section as a function of $\kappa_{hhh}$ at $\sqrt{s} = 500$  and 1000 GeV respectively. The cross section  shows  a mild sensitivity to the trilinear Higgs self-coupling. The observed dependence originates from the interplay between the diagrams involving the trilinear Higgs self-coupling and the gauge-mediated contributions. In both panels,   the cross section increases with increasing values of   $\kappa_{hhh}$. These diagrams show that gauge-mediated contributions are dominant, and the sensitivity to the trilinear Higgs self-coupling   remains mild. The  right panel of Fig.~\ref{fig3} shows   the cross section as a function of  the center-of-mass energy. In the considered energy range, the cross section increases with $\sqrt{s}$ and gradually approaches a constant value at higher energies due to the dominance of gauge-mediated contributions and the opening of the available phase space.

\begin{figure}
\begin{center}
\includegraphics[width=0.315\linewidth]{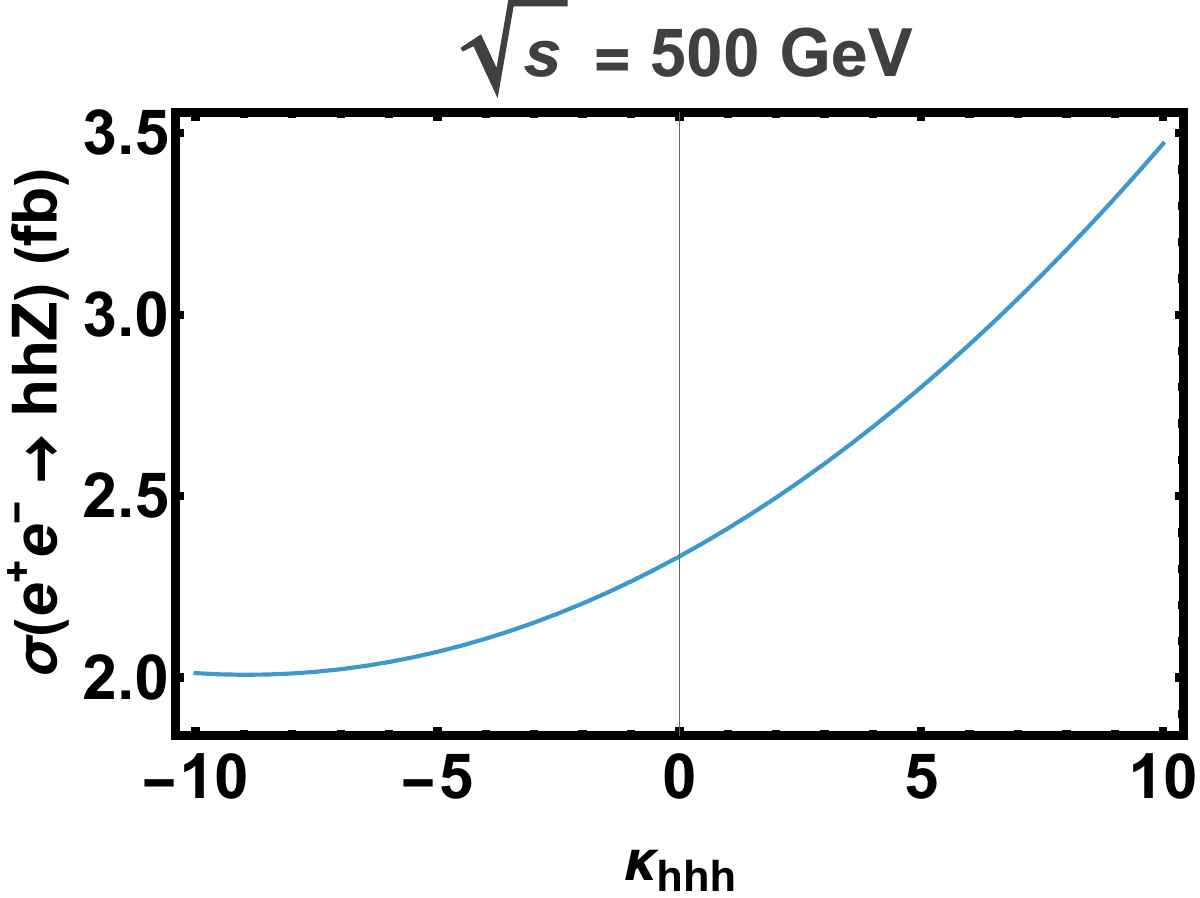}\hspace{3.24mm}
\includegraphics[width=0.3158\linewidth]{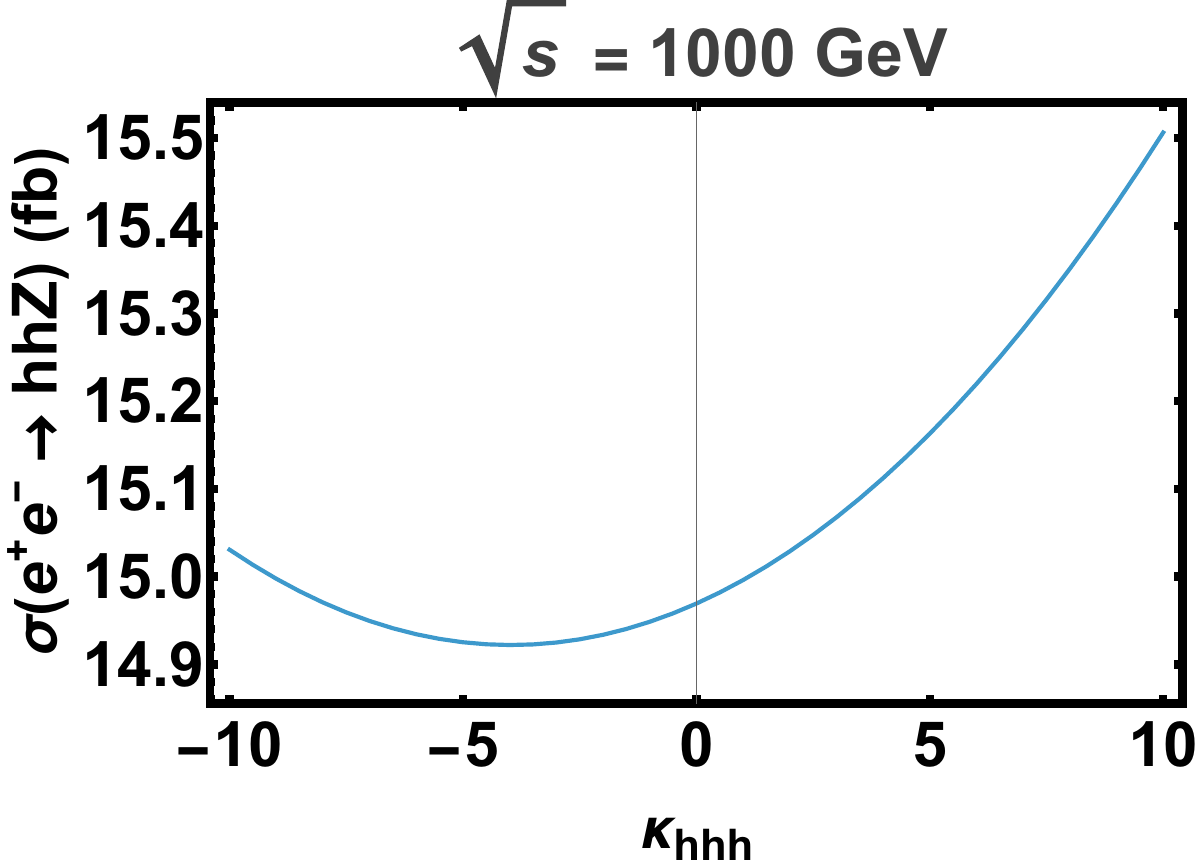}\hspace{3.24mm}    
\includegraphics[width=0.315\linewidth]{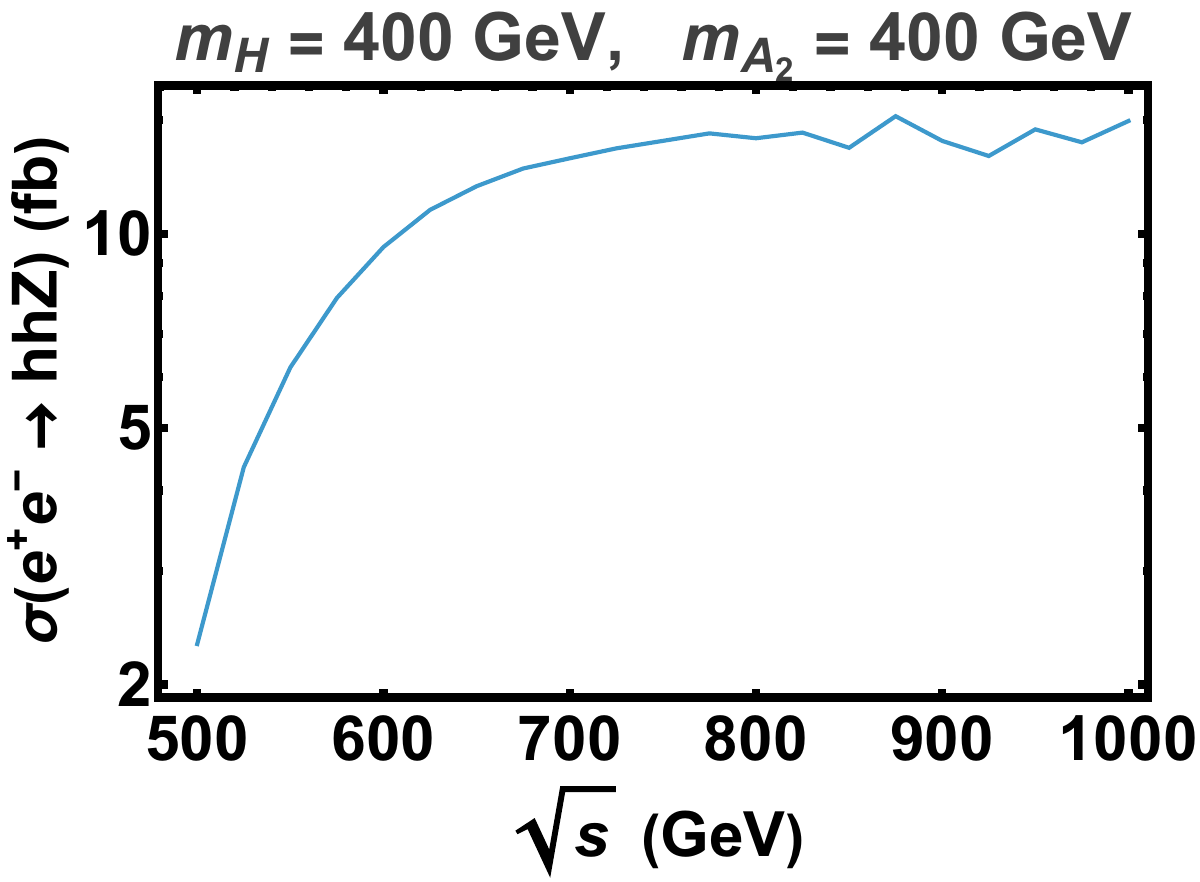}\hspace{3.24mm}
\caption{Left and middle panels: Cross sections for the process $e^+e^- \to hhZ$ as a function of the trilinear Higgs self-coupling $\kappa_{hhh}$ at $\sqrt{s}=500$ GeV  and $\sqrt{s}=1000$ GeV, respectively. Right panel: Cross section as a function of the center-of-mass energy $ \sqrt{s}$ for the process $e^+e^- \to hhZ$ .
We take $m_{h}=125$ GeV, $m_{h_0}=200$ GeV,  $ m_{A_2}=m_{H}=400 $ GeV,~ $\tan\alpha=0.4$   and $\tan\theta=1.4$.
}
\label{fig3}
\end{center}
\end{figure}

\subsubsection{Double Higgs from Vector Boson Fusion}
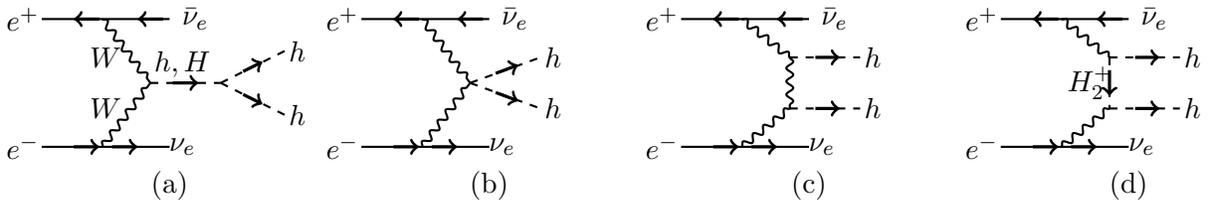
\begin{figure}[htbp]
\centering
\begin{tikzpicture}[scale=0.85, every node/.style={font=\small}]

\tikzset{
  ferm/.style={thick},
  scalar/.style={thick,dashed},
  boson/.style={thick,decorate,decoration={snake,amplitude=1.3pt,segment length=5pt}}
}

\begin{scope}[xshift=0cm]

\draw[ferm] (0,2) -- (2,2);
\draw[ferm] (0,0) -- (2,0);

\draw[->, very thick] (0.9,2) -- (0.5,2);   
\draw[->, very thick] (1.9,2) -- (1.5,2);   
\draw[->, very thick] (0.5,0) -- (0.9,0);   
\draw[->, very thick] (1.1,0) -- (1.5,0);   

\node at (-0.3,2) {$e^+$};
\node at (-0.3,0) {$e^-$};
\node at (2.4,2) {$\bar{\nu}_e$};
\node at (2.2,0) {$\nu_e$};

\draw[boson] (0.95,2) -- (1.7,1);
\draw[boson] (0.95,0) -- (1.7,1);

\node at (1.0,1.4) {$W$};
\node at (1.0,0.6) {$W$};

\draw[scalar] (1.7,1) -- (2.8,1);
\draw[very thick,->] (2.05,1) -- (2.45,1);
\node at (2.2,1.3) {$h,H$};

\draw[scalar] (2.8,1) -- (3.8,1.5);
\draw[scalar] (2.8,1) -- (3.8,0.5);

\draw[very thick,->] (3.15,1.18) -- (3.45,1.33);
\draw[very thick,->] (3.15,0.82) -- (3.45,0.67);

\node at (4.0,1.5) {$h$};
\node at (4.0,0.5) {$h$};

\node at (2,-0.6) {(a)};
\end{scope}

\begin{scope}[xshift=5cm]

\draw[ferm] (0,2) -- (2,2);
\draw[ferm] (0,0) -- (2,0);

\draw[->, very thick] (0.9,2) -- (0.5,2);
\draw[->, very thick] (1.9,2) -- (1.5,2);
\draw[->, very thick] (0.5,0) -- (0.9,0);
\draw[->, very thick] (1.1,0) -- (1.5,0);

\node at (-0.3,2) {$e^+$};
\node at (-0.3,0) {$e^-$};
\node at (2.4,2) {$\bar{\nu}_e$};
\node at (2.2,0) {$\nu_e$};

\draw[boson] (0.95,2) -- (1.7,1);
\draw[boson] (0.95,0) -- (1.7,1);

\draw[scalar] (1.7,1) -- (2.8,1.4);
\draw[scalar] (1.7,1) -- (2.8,0.6);

\draw[very thick,->] (2.10,1.15) -- (2.45,1.28);
\draw[very thick,->] (2.10,0.85) -- (2.45,0.72);

\node at (3.0,1.4) {$h$};
\node at (3.0,0.6) {$h$};

\node at (2,-0.6) {(b)};
\end{scope}

\begin{scope}[xshift=10cm]

\draw[ferm] (0,2) -- (2,2);
\draw[ferm] (0,0) -- (2,0);

\draw[->, very thick] (0.9,2) -- (0.5,2);
\draw[->, very thick] (1.9,2) -- (1.5,2);
\draw[->, very thick] (0.5,0) -- (0.9,0);
\draw[->, very thick] (1.1,0) -- (1.5,0);

\node at (-0.3,2) {$e^+$};
\node at (-0.3,0) {$e^-$};
\node at (2.4,2) {$\bar{\nu}_e$};
\node at (2.2,0) {$\nu_e$};

\draw[boson] (0.95,2) -- (1.7,1.4);
\draw[boson] (0.95,0) -- (1.7,0.6);
\draw[boson] (1.7,1.4) -- (1.7,0.6);

\draw[scalar] (1.7,1.4) -- (2.8,1.4);
\draw[scalar] (1.7,0.6) -- (2.8,0.6);

\draw[very thick,->] (2.10,1.4) -- (2.48,1.4);
\draw[very thick,->] (2.10,0.6) -- (2.48,0.6);

\node at (3.0,1.4) {$h$};
\node at (3.0,0.6) {$h$};

\node at (2,-0.6) {(c)};
\end{scope}

\begin{scope}[xshift=15cm]

\draw[ferm] (0,2) -- (2,2);
\draw[ferm] (0,0) -- (2,0);

\draw[->, very thick] (0.9,2) -- (0.5,2);
\draw[->, very thick] (1.9,2) -- (1.5,2);
\draw[->, very thick] (0.5,0) -- (0.9,0);
\draw[->, very thick] (1.1,0) -- (1.5,0);

\node at (-0.3,2) {$e^+$};
\node at (-0.3,0) {$e^-$};
\node at (2.4,2) {$\bar{\nu}_e$};
\node at (2.2,0) {$\nu_e$};

\draw[boson] (0.95,2) -- (1.7,1.4);
\draw[boson] (0.95,0) -- (1.7,0.6);

\draw[scalar] (1.7,1.4) -- (1.7,0.6);
\draw[very thick,->] (1.7,1.2) -- (1.7,0.8);
\node at (1.4,1.0) {$H_2^+$};

\draw[scalar] (1.7,1.4) -- (2.8,1.4);
\draw[scalar] (1.7,0.6) -- (2.8,0.6);

\draw[very thick,->] (2.10,1.4) -- (2.48,1.4);
\draw[very thick,->] (2.10,0.6) -- (2.48,0.6);

\node at (3.0,1.4) {$h$};
\node at (3.0,0.6) {$h$};

\node at (2,-0.6) {(d)};
\end{scope}

\end{tikzpicture}
\caption{Feynman diagrams for $ e^+ e^- \rightarrow hh \nu_e \bar{\nu}_e $.} \label{fig:vvhh_feynman}
\end{figure}
The double Higgs production process $ e^+ e^- \rightarrow  h h\nu_{e} \bar{\nu}_{e}$ is determined within  the Effective-W-Boson Approximation (EWA), which is widely used in studies of vector-boson-fusion processes~\cite{djouadi2,djouadi,osland, Li:2017daq, Arhrib:2015gra}.  
The EWA approximation is expected to provide more accurate results in the kinematic region where  $\sqrt{s}\gg m_W$. Comparisons between EWA and exact matrix-element calculations available in the literature show that, at relatively low center-of-mass energies, the EWA may overestimate the magnitude of the cross section. For example, Ref.~\cite{ djouadi} found deviations reaching up to a factor of two for small  masses and/or low center-of-mass energies. Therefore, the results presented at $\sqrt{s}=500 $ GeV should be considered as an approximate estimate of the VBF production rate. Nevertheless, the EWA remains a valuable tool for studying the general trend of the process and the enhancement patterns generated by the extended scalar sector.

Following Refs.~\cite{djouadi2,osland, Li:2017daq}, the total cross section can be written as

\begin{equation}
\sigma(e^+ e^- \rightarrow hh \nu_e \bar{\nu_e})
= \int_{x_{\min}}^{1} dx \; \frac{dL}{dx} \; \hat{\sigma}_{WW}(x), \label{diff}
\end{equation}
where $x_{\min} = 4r_{h} = 4m_{h}^2/s$.
The differential luminosity function is described by \cite{osland}
\begin{equation}
\frac{dL(x)}{dx} =
\frac{G_F^2 m_W^4}{8\pi^4} \frac{1}{x}
\left[-(1+x)\ln x - 2(1-x)\right].
\end{equation}

In particular, the subprocess cross section  can be expressed as \cite{osland}
\begin{equation}
\hat{\sigma}_{WW}(x) =
\frac{G_F^2 \hat{s}}{64\pi}
\left[
4\beta_{h}
\left(
\frac{3r_{h}\,\kappa_{hVV}\,\kappa_{hhh}}{1-r_{h}}
+
\frac{3r_{h}\,\kappa_{HVV}\,\kappa_{Hhh}}{1-r_{H}}
+
\kappa_{hhVV}
\right)^2
\right. \label{eq4}
\end{equation}

\begin{equation*}
+\,2
\left(
\frac{3r_{h}\,\kappa_{hVV}\,\kappa_{hhh}}{1-r_{h}}
+
\frac{3r_{h}\,\kappa_{HVV}\,\kappa_{Hhh}}{1-r_{H}}
+
\kappa_{hhVV}
\right)
\left(
\kappa_{hVV}^2 F_1
+
\kappa_{H_2^\pm h W}^2 F_2
\right)
\end{equation*}

\begin{equation*}
\left.
+\,\beta_{h}^{-1}
\left(
\kappa_{hVV}^4 F_3
+
\kappa_{H_2^\pm h W}^4 F_4
+
4\,\kappa_{hVV}^2 \kappa_{H_2^\pm hW}^2 F_5
\right)
\right],
\end{equation*}

where $\beta_{h} = \sqrt{1 - 4r_{h}}$. The functions $F_i$ $(i = 1,\dots,5)$ are listed in Appendix~\ref{app:VBF}.  The  parameters $r_i$ $(i = h, Z, H, H_{2}^{\pm})$ are defined in terms of $\hat{s} = xs$, for instance $r_H = m_{H}^2 / \hat{s}$. In our numerical calculations, we use the analytical expressions given above.

\begin{figure}[t]
\begin{center}
\includegraphics[width=0.42\linewidth]{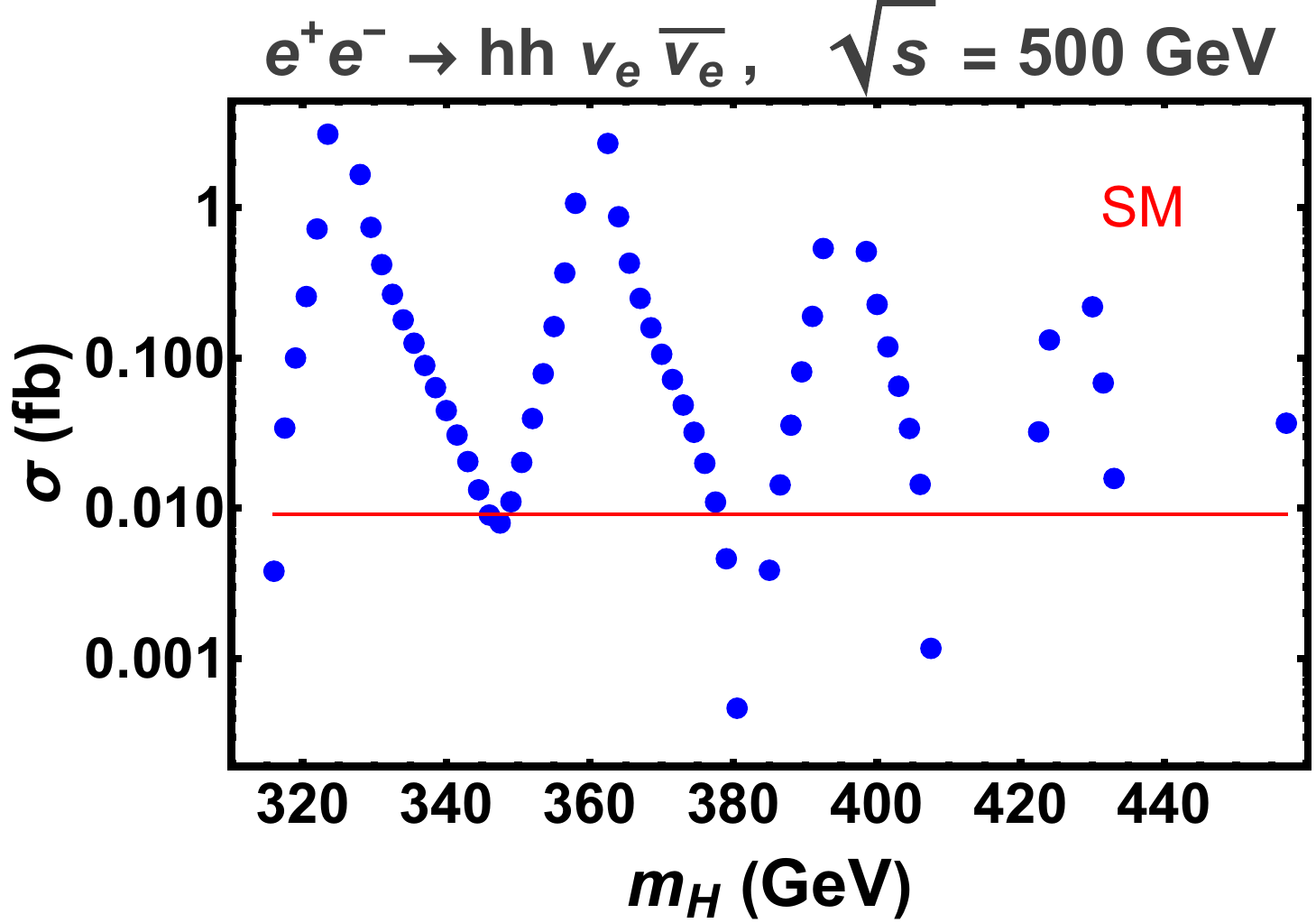}\hspace{3.25mm}
\includegraphics[width=0.42\linewidth]{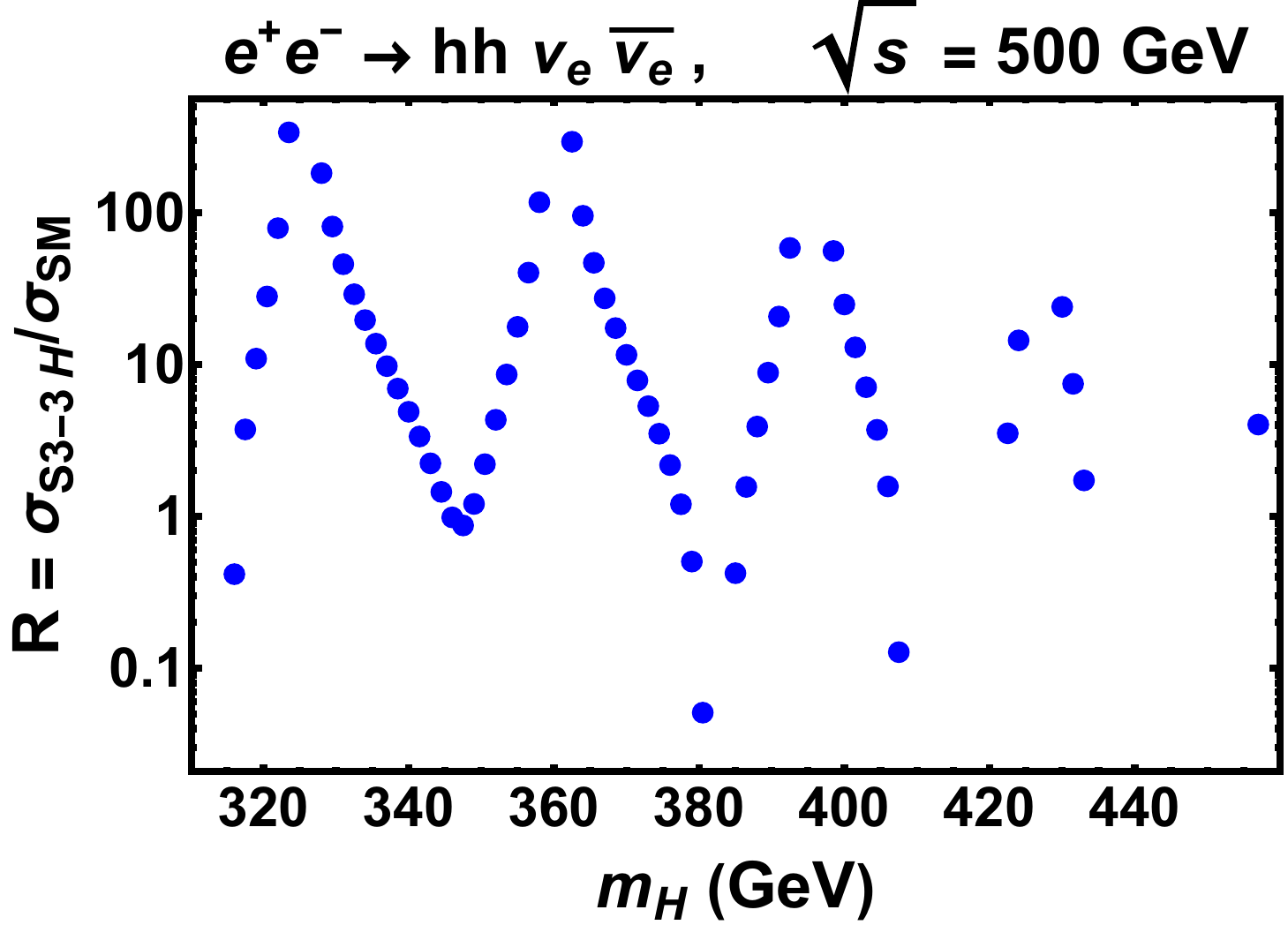}\hspace{3.25mm}

\caption{Left panel: Cross section for the  process $ e^+ e^- \rightarrow  h h\nu_{e} \bar{\nu}_{e}$  as a function of $m_{H}$. Right panel: The  ratio $R=\sigma_{\text{S3-3H}}/\sigma_{\text{SM}}$ as a function of $m_{H}$.  Input parameters: $m_{h}=125$ GeV,  $m_{h_0}=200$ GeV,   $ m_{H_{2}^{\pm}}=350$ GeV, $\tan\alpha=0.4$ and $\tan\theta=1.4$.}
\label{fig4}
\end{center}
\end{figure}

\begin{figure}[t]
\begin{center}

\includegraphics[width=0.42\linewidth]{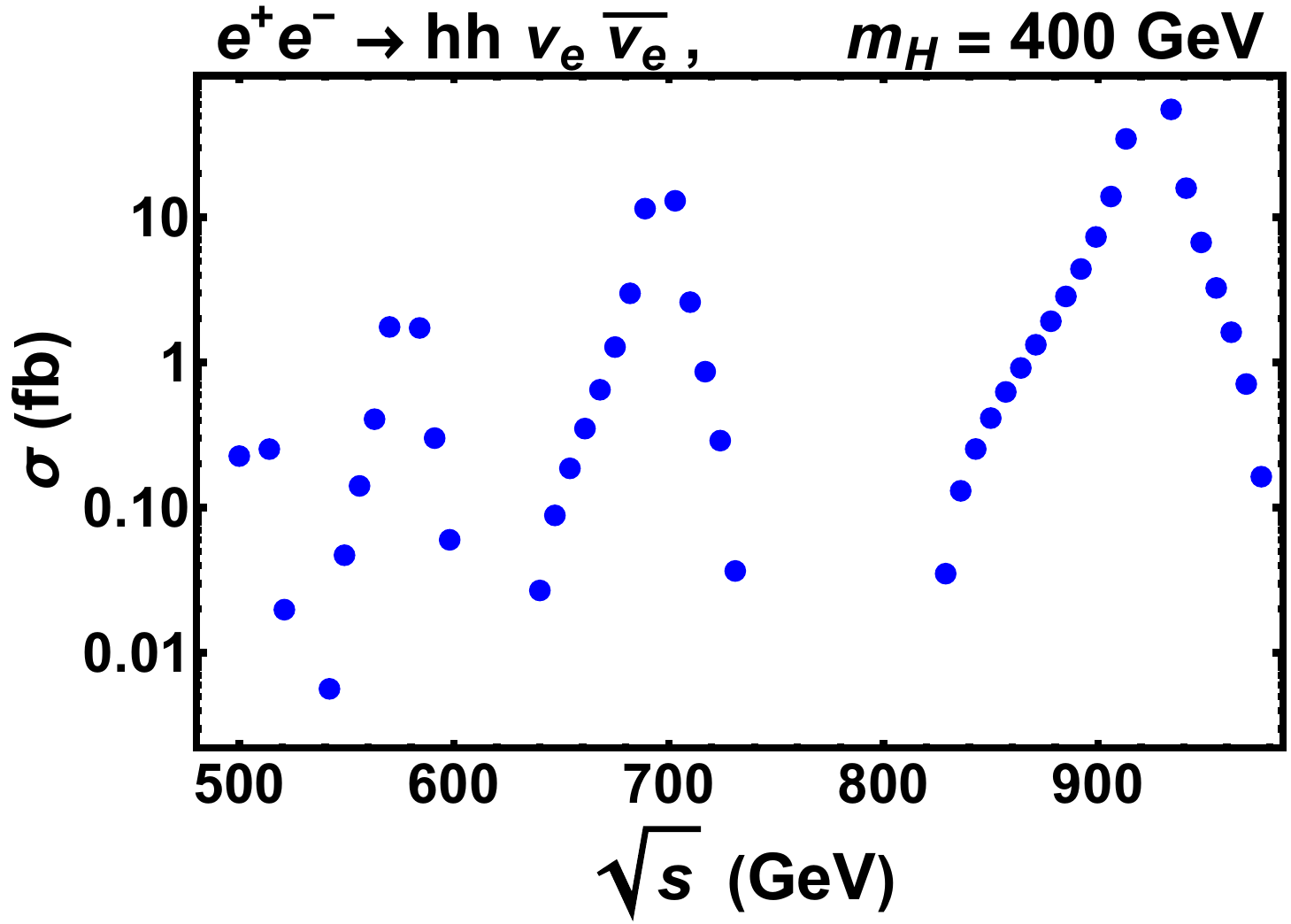}\hspace{3.25mm} 
\includegraphics[width=0.42\linewidth]{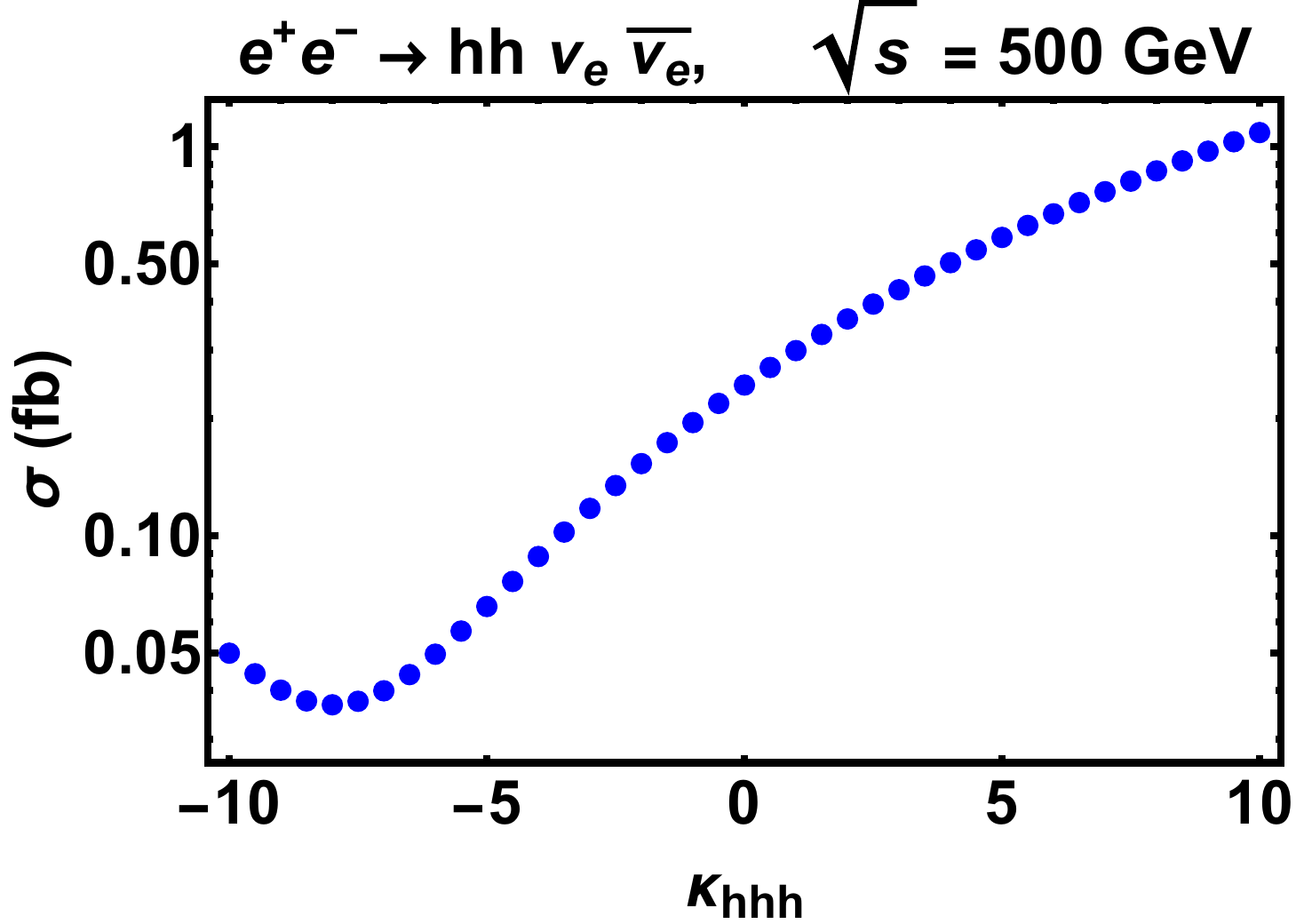}\hspace{3.25mm}\\    \vspace{5mm}
\caption{
Cross sections for the process $ e^+ e^- \rightarrow  h h\nu_{e} \bar{\nu}_{e}$.   Left panel: Cross section as a function of the center-of-mass energy $\sqrt{s}$. Right panel: Cross section as a function of the trilinear Higgs self-coupling $\kappa_{hhh}$. 
Input parameters: $m_{h}=125$ GeV, $m_{h_0}=200$ GeV, $m_{H_2^\pm}=350$ GeV, $m_{H}=400$ GeV,  $\tan\alpha=0.4$, and $\tan\theta=1.4$.
}
\label{fig8}
\end{center}
\end{figure}

The results for the  production rate $ e^+ e^- \rightarrow  h h\nu_{e} \bar{\nu}_{e}$ are shown  in Fig.~\ref{fig4}. In the left panel of Fig.~\ref{fig4}, one can see that the cross section shows a non-linear dependence on the  heavy scalar mass $m_{H}$. Several regions illustrate enhancements above the SM expectation (indicated by the horizontal line). These enhancements emerge from scalar contributions and their constructive interference with gauge interactions. In the right panel, we present the ratio $R = \sigma_{S3-3H}/\sigma_{SM}$ as a function of  $m_{H}$ at $\sqrt{s}=500$ GeV. It can be seen from this figure that WW-fusion production of a $ h $ pair in the S3-3H model  is significantly larger than the SM prediction. The relative deviation $\Delta R $ varies approximately  in the range 
$-0.95 \lesssim \Delta R \lesssim \mathcal{O}(10^{2})$ for $\sqrt{s}=500$ GeV and $310~\mathrm{GeV} \leq m_H \leq 560~\mathrm{GeV}$.

 The left panel of Fig.~\ref{fig8} displays the production rate as a function of the center-of-mass energy for a fixed heavy scalar mass. The cross section increases with energy, reflecting the growing importance of VBF  in the process $e^+e^- \to hh \nu_e \bar{\nu}_e$ at higher energies. The left panel of Fig.~\ref{fig8} illustrates that the general features remain stable as the center-of-mass energy increases toward the TeV regime, where the validity of the EWA is expected to improve. The right panel of Fig.~\ref{fig8} shows the   cross section   as a function of the trilinear Higgs self-coupling modifier $\kappa_{hhh}$ at $\sqrt{s}=500$ GeV. The cross section has a characteristic non-linear behavior.  The cross section is  minimum at negative values of $\kappa_{hhh}$ and increases for both larger negative and positive values.  This behavior can be explained by the interplay between the self-coupling contribution and other production diagrams, leading to constructive or destructive interference depending on the value of $\kappa_{hhh}$.   In this analysis, we consider  $\kappa_{hhh}$ as an independent parameter and study the sensitivity of the $e^+e^- \to hh \nu_e \bar{\nu}_e$ process to variations of the trilinear Higgs self-coupling, while keeping the other couplings fixed in this plot. 

\begin{figure}[t]
\begin{center}
\includegraphics[width=0.46\linewidth]{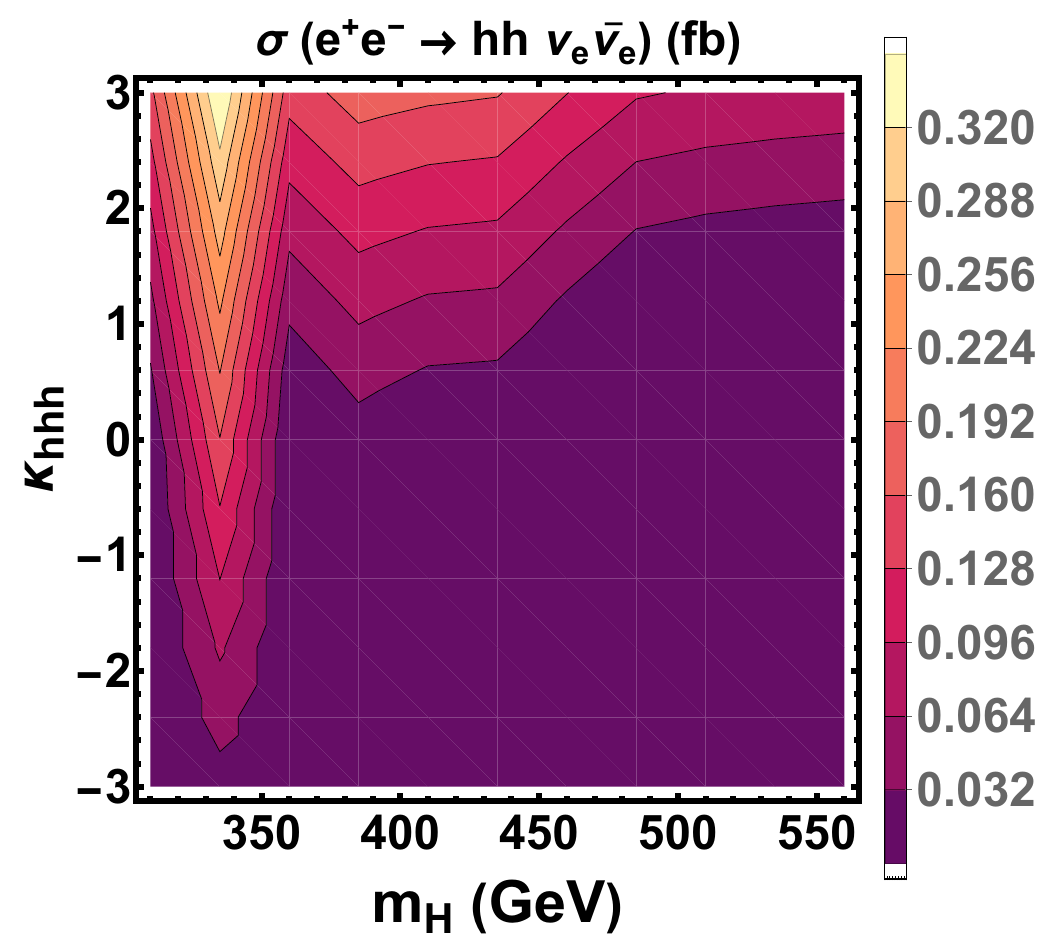}\hspace{3.25mm}
\includegraphics[width=0.44\linewidth]{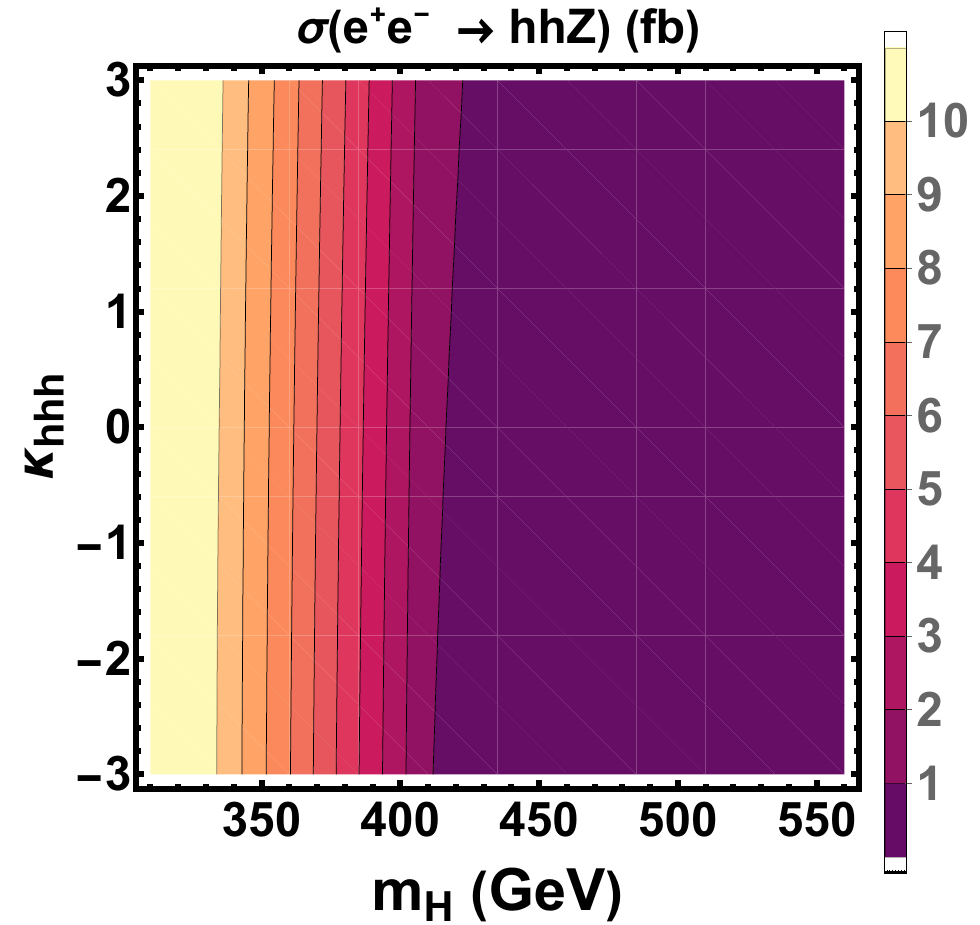}\hspace{3.25mm}\\    \vspace{5mm}
\
\caption{The contour plot of $\sigma( e^+ e^- \rightarrow  h h\nu_{e} \bar{\nu}_{e})$(left) and $\sigma(e^+e^- \to hhZ)$ (right ) in the plane $ m_{H} $ versus $ \kappa_{hhh} $.
We take $m_{h}=125$ GeV, $m_{h_0}=200$ GeV,  $ m_{A_2}=400 $ GeV~ $m_{H_2^\pm}=350$ GeV, $\sqrt{s}=500$ GeV, $\tan\alpha=0.4$  and $\tan\theta=1.4$.}
\label{fig5}
\end{center}
\end{figure}

In Fig.~\ref{fig5}, we indicate the behavior of double Higgs production via the VBF and HS channels in  the $ (m_{H}, \kappa_{hhh}) $ plane at $ \sqrt{s}=500 $ GeV. Two production mechanisms show different sensitivities to the trilinear Higgs-self coupling.  The VBF channel shows a more pronounced dependence on $\kappa_{hhh},$ while the HS channel presents a comparatively weaker dependence. A quantitative comparison shows that, over the considered $(\kappa_{hhh})$ range, the HS cross section varies from 2.15 fb to 2.59 fb, corresponding to $\sigma_{\max}/\sigma_{\min}=1.20.$ In contrast, the VBF cross section varies from 0.12 fb to 0.43 fb, corresponding to $\sigma_{\max}/\sigma_{\min}=3.65.$ These two channels show that the VBF channel provides a stronger dependence on the trilinear Higgs self-coupling in the allowed region of parameter space.

In conclusion,  double Higgs production  is a powerful tool to test the SM and search for new physics through the  predicted  cross sections in  the HS  and VBF  production modes. The cross section of both processes increases with center-of-mass energy, thus the processes gain importance at  higher energies  and they provide a strong motivation to explore  higher center-of-mass energies.

In conclusion, double Higgs production provides a useful framework for studying possible deviations from the SM through the predicted cross sections in the HS and VBF channels.

\section{CONCLUSIONS}

In this study, we have considered   double SM-like Higgs ($h$) production at future $ e^{+}e^{-} $ colliders in the  S3-3H model  without CP violation. We impose theoretical constraints such as perturbative unitarity and vacuum stability  and experimental constraints from Higgs signal strength measurements and direct searches, via HiggsBounds and HiggsSignals tools  to test the viability of the model. Therefore,  we focused on  the allowed region of the parameter space in our analysis of double Higgs production via $e^+ e^- \to hhZ$ and $e^+ e^- \to hh \nu_e \bar{\nu}_e$ modes.

In this work, we have first analyzed double Higgs production via the Higgs-strahlung process $e^+e^- \to hhZ$. The production process $e^+ e^- \to hhZ$ can exhibit  large deviations from the SM prediction, with the ratio $R = \sigma^{\text{S3-3H}}/\sigma^{\text{SM}}$ exceeding unity by up to a few orders of magnitude in certain regions of the parameter space. These results illustrate that double Higgs production can serve as  a sensitive probe of extended Higgs sectors, as  large deviations from the SM can arise from  enhanced contributions of additional scalar states. Our calculation shows that the enhancement of the
$e^{+}e^{-}\to hhZ$ cross section is dominated by resonant $ A_2 $ exchange, while interference effects remain small and slightly destructive throughout the allowed parameter space.

We also investigated Vector Boson Fusion  via $e^+ e^- \to hh \nu_e \bar{\nu}_e$. The cross section can  strongly increase due to scalar contributions and their constructive interference with gauge-mediated diagrams, with the ratio $R = \sigma^{\text{S3-3H}}/\sigma^{\text{SM}}$ reaching values well above unity. The production rate becomes larger with the center-of-mass energy, showing that the VBF becomes important at higher energies. In addition, the  dependence  on the trilinear Higgs self-coupling  in the VBF channel provides a sensitive probe of both the extended scalar sector and the Higgs self-coupling at future $e^+e^-$ colliders. In comparison with  the HS process, the VBF channel exhibits a relatively stronger sensitivity to the trilinear Higgs self-coupling, thus it becomes a potentially more suitable channel for testing the Higgs self-interaction at high energies.

To conclude, this work has investigated the impact of the trilinear Higgs self-coupling and extra Higgs bosons ($ H $ and $ A_{2} $) on double Higgs production at future $e^+e^-$ colliders such as the International Linear Collider (ILC)~\cite{86,89} and the Compact Linear Collider (CLIC)~\cite{87,88}. Moreover, studying double Higgs production at various $\sqrt{s}$ values may help distinguish the S3-3H model from alternative new physics scenarios. It  can also  be used to probe the viability  of the S3-3H model in nature. However, a reliable assessment of the observability of the predicted cross-section enhancements requires dedicated background and detector-level studies, which are beyond the scope of the present work.

\noindent

\appendix
\section{Coefficients for the Process $ e^+e^- \rightarrow hhZ $ }
\label{app:A}
In this appendix, we present the explicit expressions for the coefficients  entering Eq.~(\ref{eq3}) as follows:

\begin{align}
f_a &= x_3^2 + 8 r_Z, \\
f_b &= (x_1^2 - 4 r_h)\left[(y_1 - r_Z)^2 - 4 r_Z r_h \right], \\
g_{ab} &= r_Z \left[ 2(r_Z - 4 r_h) + x_1^2 + x_2(x_2 + x_3) \right] - y_1(2y_2 - x_1 x_3), \\
g_{bb} &= (y_3 - x_1 x_2 - x_3 r_Z - 4 r_h r_Z)(2y_3 - x_1 x_2 - 4 r_h + 4 r_Z) \nonumber \\
&\quad + r_Z^2 (4 r_h + 6 - x_1 x_2) + 2 r_Z (r_Z^2 + y_3 - 4 r_h).
\end{align}

\section{Coefficients for the Process $  e^+ e^- \rightarrow  h h\nu_{e} \bar{\nu}_{e}$}
\label{app:VBF}

In this appendix, we list the coefficient functions $F_i$ $(i=1,\dots,5)$ appearing in Eq.~(\ref{eq4}) \cite{osland}.

\begin{align}
F_1 &= 8\left[2r_W + (r_{h}-r_W)^2\right] l_W
      - 4\beta_{h}\left(1+2r_{h}-2r_W\right), \\
F_2 &= 8(r_{h}-r_C)^2 l_C
      - 4\beta_{h}\left(1+2r_{h}-2r_W\right), \\
F_3 &= 8\beta_{h}\left[2r_W + (r_{h}-r_W)^2\right]
      \left[2r_W + 1 - 3(r_{h}-r_W)^2\right]\frac{l_W}{a_W}
      \nonumber \\
    &\quad + 16\left[2r_W + (r_{h}-r_W)^2\right]^2 y_W
      + 16\beta_{h}^2(1+a_W)^2, \\
F_4 &= 8\beta_{h}(r_{h}-r_C)^2
      \left[1 - 3(r_{h}-r_C)^2\right]\frac{l_C}{a_C}
      \nonumber \\
    &\quad + 16(r_{h}-r_C)^4 y_C
      + 16\beta_{h}^2(1+a_C)^2, \\
F_5 &= \frac{\beta_{h}}{4}(z_W l_W + z_C l_C)
      + 8\beta_{h}^2(1+a_W)(1+a_C).
\end{align}

\begin{align}
l_{W,C} &=
\ln\left(
\frac{1-2r_{h}+2r_{W,C}-\beta_{h}}
     {1-2r_{h}+2r_{W,C}+\beta_{h}}
\right), \\
y_{W,C} &=
\frac{2\beta_{h}^2}
{\left(1-2r_{h}+2r_{W,C}\right)^2-\beta_{h}^2}, \\
a_{W,C} &= -\frac{1}{2} + r_{h} - r_{W,C}.
\end{align}

where the functions $z_W$ and $z_C$ are given by
\begin{align}
z_W &=
\frac{(1+2a_W)^2}{a_C-a_W}\left[8r_W + (1+2a_W)^2\right]
+ \frac{(1-2a_W)^2}{a_C+a_W}\left[8r_W + (1+2a_W)^2\right], \\
z_C &=
-\frac{(1+2a_C)^2}{a_C-a_W}\left[8r_W + (1+2a_C)^2\right]
+ \frac{(1+2a_C)^2}{a_C+a_W}\left[8r_W + (1-2a_C)^2\right].
\end{align}
We define the charged scalar  $H_2^\pm$ contribution as
\begin{equation}
r_C = \frac{m_{H_2^\pm}^2}{\hat{s}},
\end{equation}
such that the shorthand notation $r_{W,C}$ represents either $r_W$ or $r_C$, depending on the relevant contribution.

\end{document}